\documentclass[apj]{emulateapj}%[10pt,preprint]{aastex}
\usepackage{graphicx}
\usepackage{natbib}
\usepackage{amsmath}
\usepackage{apjfonts}

\newcommand{\Msun}{M_{\sun}}

\begin{document}

\title{No flares from Gamma-Ray Burst afterglow blast waves encountering sudden circumburst density change}
\author{Ilana Gat$^1$, Hendrik van Eerten$^1$, Andrew MacFadyen$^1$}
\affil{
  $^1$ Center for Cosmology and Particle Physics, Physics Department, New York University, New York, NY 10003\\
}
\begin{abstract}

Afterglows of gamma-ray bursts are observed to produce light curves with the flux following power law evolution in time. However, recent observations reveal bright flares at times on the order of minutes to days. One proposed explanation for these flares is the interaction of a relativistic blast wave with a circumburst density transition. In this paper, we model this type of interaction computationally in one and two dimensions, using a relativistic hydrodynamics code with adaptive mesh refinement called \textsc{ram}, and analytically in one dimension. We simulate a blast wave traveling in a stellar wind environment that encounters a sudden change in density, followed by a homogeneous medium, and compute the observed radiation using a synchrotron model. We show that flares are not observable for an encounter with a sudden density increase, such as a wind termination shock, nor for an encounter with a sudden density decrease. Furthermore, by extending our analysis to two dimensions, we are able to resolve the spreading, collimation, and edge effects of the blast wave as it encounters the change in circumburst medium. In all cases considered in this paper, we find that a flare will not be observed for any of the density changes studied.
\end{abstract}

\section{Introduction}\label{sec:intro}

Since the discovery of gamma-ray bursts (GRBs), there has been increasing interest in their properties and behavioral characteristics. Currently, GRBs are thought to be the result of either a massive star collapsing \citep{woosley1993, macfadyen1999} or compact binary systems merging \citep{eichler1989}, launching a collimated relativistic blast wave into the circumburst medium. The blast wave sweeps up, shocks, and accelerates the circumburst electrons as the circumburst medium slows the blast wave itself. The shock-accelerated electrons produce synchrotron radiation as they interact with small scale magnetic fields behind the shock front (e.g.,~\citealt{Paczynski1993, Katz1994, Sari1996, Wijers1997, Wijers1999, Panaitescu2002}). This radiation creates an afterglow signal that can be observed for days at X-ray and optical frequencies and for even longer time scales at radio frequencies.

Much research has been done to understand the blast wave that causes the observed afterglow signal, and the evolution of the ultra-relativistic early time and non-relativistic late time phases of the blast wave are known: they can be described by the self-similar Blandford-McKee (BM:~\citealt{Blandford1976})  and the Sedov-Von Neumann-Taylor (ST:~\citealt{Sedov1959, vonNeumann1961, Taylor1950}) solutions respectively. Such analytical solutions describe the radial outflow of a collimated blast wave at early times, as well as the spherical outflow of the blast wave at late times. Yet there are currently no exact analytical solutions for the intermediate phases of the evolution, although there are approximations (e.g.,~\citealt{Peer2012, Huang1999} for the spherical case), describing the dynamics of the deceleration and spreading of the blast wave. The spreading of the blast wave as it decollimates has been treated analytically \citep{ rhoads1999, granot2012} and recently modeled computationally \citep{Zhang2009,  Wygoda2011, vanEerten2012observationalimplications, vanEerten2012boxfit, DeColle2012}.

Along with studying the evolution of the blast wave itself, much research has been devoted to studying the shape of the emitted light curve. The general model of the afterglow light curve is a smooth curve with the slope being a function of the density of the surrounding medium as well as the power-law slope of the distribution of the accelerated electron population at the shock front. Recent observations, though, have shown this model is not always sufficient. Flares in X-ray afterglows were first detected with {\it BeppoSAX} \citep{Frontera2000} and once {\it Swift} was launched in 2004, it became clear that afterglow flares \citep{burrows2005, nousek2006, obrien2006} as well as optical variability in the early stages of the burst \citep{stanek2007} were a common occurrence. 

To explain the causes of these flares, researchers have began to study the interaction of a blast wave with complex structures such as wind termination shocks \citep{Dai2002, Ramirez-Ruiz2005,Peer2006, Eldridge2006, Nakar2007}, clumps \citep{Ramirez-Ruiz2005}, magnetic shocks \citep{Yost2003}, collisions with nearby star environments \citep{Mimica2011}, and massive shells \citep{Mesler2012}. Others have speculated the flares could be caused by slower ejecta catching up with the blast wave, reenergizing it at later times (e.g.,~\citealt{Rees1998, Kumar2000, Sari2000}), from delayed magnetic dissipation~\citep{Giannos2006}, or from magnetic regulation of the accretion flow~\citep{Proga2006}. Alternatively, it has been theorized that flaring is a result of late time engine activity~\citep{Falcone2007, Perna2010, Margutti2011} which, for example, is from multiple shells being ejected~\citep{Maxham2009, Vlasis2011}, mass influx~\citep{Metzger2008, Lee2009}, or fragmentation in the collapsing stellar core~\citep{King2005} or accretion disk~\citep{Perna2006, Masada2007}. There are many hypotheses of the cause of the rebrightening in the light curve (see also~\citealt{Ioka2005}), and in this paper, we discuss the hypothesis of a transition in circumburst medium from a stellar wind to a homogeneous medium (e.g., an interstellar medium (ISM)), including wind termination shocks as well as sudden lower density regions. These types of density changes have been numerically studied in much detail and are from the formation of a wind reverse shock (e.g.,~\citealt{vanMarle2006, Eldridge2006}).

Pe'er and Wijers (PW:~\citealt{Peer2006}) theorized that the flares were the result of the blast wave radiation interfering with the reverse shock of the stellar wind, causing a transition period in the light curve. However, Nakar and Granot (NG:~\citealt{Nakar2007}) were unable to reproduce PW's results using one dimensional simulations and argued that the flares are not caused by a wind termination shock nor a density jump in a uniform external medium. NG concluded that the origin of the flares in afterglow light curves has still yet to be discovered. \cite{vanEerten2009windtermination} reconciled this discrepancy by explaining NG was correct in stating a wind termination shock does not cause rebrightening, although the results of PW correctly follow from their model assumptions. \cite{Mesler2012} revisited this idea of a wind termination shock but instead of one shock, added higher density shells that can realistically be expected to occur, and claimed that flares were observed as the blast wave encountered these higher density shells. 

In this paper, we address a number of circumburst medium interaction scenarios not previously explored. We also extend the analysis from one dimension to two dimensions to understand the effects of spreading of a collimated flow at the shock front which was not previously considered. Specifically, we address the following questions in our analysis using high resolution hydrodynamic simulations in one and two dimensions.
\begin{itemize}
\item When a collimated blast wave traveling in a stellar wind environment encounters a wind termination shock, how does the size of the density jump affect the dynamics and resulting light curve?
%\item Does the Lorentz factor of the fluid behind the shock at the time of the encounter affect the dynamics and resulting light curve?
\item What happens when there is a density drop instead of a jump? Does the blast wave speed up, and in turn, recollimate? Will this cause a rebrightening or flare in the light curve?
\item When the blast wave encounters an extreme density increase, does it immediately spread outward from this high energy collision and cause flares in the light curve? Does this sideways spreading depend on the size of the density jump?
\end{itemize}

To answer these questions, we use a numerical relativistic hydrodynamics (RHD) code with adaptive mesh refinement called \textsc{ram} \citep{Zhang2006, Zhang2009} for our numerical simulations. We use the radiation calculation methodology explained in~\cite{vanEerten2010offaxis, vanEerten2010transrelativistic} to calculate synchrotron afterglow light curves and spectra. In Section~\ref{sec:D} we discuss the dynamics of the simulations which is followed by Section~\ref{sec:LC} discussing the resulting light curves. We summarize and conclude our findings in Section~\ref{sec:SaC}. Details of the resolution of our simulations are explained in Appendix~\ref{App:AppendixA}, and we discuss the specialized case of a spike in circumburst medium in Appendix~\ref{App:AppendixB}.

\section{Dynamics of Blast Wave Encounters}\label{sec:D}

A key aspect in understanding the role of the circumburst medium in the flux emitted from a blast wave is the dynamics of the blast wave in that medium. In this section, we discuss those dynamics.

\subsection{Initial Conditions}\label{sec:IC}

The initial conditions for an adiabatic blast wave (``jet'') formed by a GRB are described by the BM solution in spherical coordinates. For this paper, we use a conic section truncated at a certain opening angle for the initial setup, instead of the full spherical solution for, at early times, spreading has not yet occurred. Using the initial conditions for a jet flowing in the radial direction with an opening angle, $\theta_0$, and a jet energy, $E_{jet}$, we obtain the isotropic equivalent energy, $E_{iso}$.
\begin{equation}
E_{jet} = E_{iso}(1 - \cos\theta_0)\approx E_{iso}\theta_0^2/2
\end{equation}

The initial radius, time, and dynamics of the interaction of the jet with the circumburst medium are calculated using the BM self similar solution. For this paper, we represent the circumburst medium interior to the density change using a power-law density profile which is described as:
\begin{equation}\label{eq:external_density}
\rho_{ext} = \rho_{ref}\left(\frac{r}{R_{ref}}\right)^{-k}
\end{equation}
where $\rho_{ref}$ is the circumburst density at a reference radius, $R_{ref}$, and $k=2\,\, \mathrm{or}\,\, 0$ to represent the stellar wind or ISM environment respectively. An ISM is a region of constant density, and throughout this paper we use ``ISM'' and ``homogeneous medium'' interchangeably.  The isotropic energy of the system is given by (BM):
\begin{equation}\label{eq:Eiso}
E_{iso} = \frac{8\pi\rho_{ext}\gamma^2 t^3 c^5}{17 - 4k}
\end{equation}
where $\gamma$ is the fluid  Lorentz factor just behind the shock, and $\Gamma = \sqrt{2}\gamma$ is the Lorentz factor of the shock itself. Substituting Equation~\ref{eq:external_density} into Equation~\ref{eq:Eiso}, and using the approximation that at ultra-relativistic speeds, $R_0 \approx cT_0$, the initial radius of the shock is:
\begin{equation}\label{eq:r0}
R_0 = \left(\frac{(17-4k)E_{iso}}{8\pi \rho_0 R_{ref}^k c^2 \gamma_0^2}\right)^{\frac{1}{3-k}}.
\end{equation}
Using these expressions, we can solve for the initial time of the simulation.
\begin{equation}\label{eq:initial_time}
T_0 = \frac{R_0}{c}\left( 1 + \frac{1}{2(4-k)\Gamma_0^2}\right)
\end{equation}

As the jet propagates into the circumburst medium described by Equation~\ref{eq:external_density}, the Lorentz factor and radius of the shock over time are well known--the radius will follow a similar solution to Equation~\ref{eq:initial_time}. The subscript of ``FS'' is used to denote the front of the shock, or the forward shock:
\begin{equation}\label{eq:renc}
R_{FS} = cT\left(1- \frac{1}{2(4-k)\Gamma^2}\right).
\end{equation}
The Lorentz factor evolves as
\begin{equation}\label{eq:gamma_enc}
\Gamma = \Gamma_0\left(\frac{T}{T_0}\right)^{-\frac{3-k}{2}}.
\end{equation}
The initial conditions at $T_0$ of our numerical simulations use the full BM profile, but for simplicity in our analytical solution, we assume the blast wave is a homogeneous shell of constant density within the shell, and the back of the shell being at a radius of $R_{back}$:
\begin{equation}\label{eq:rback}
R_{back} = R_{FS}\left(1 - \frac{1}{2(3-k)\Gamma_{FS}^2}\right).
\end{equation}
Equation~\ref{eq:rback} is derived to the leading order of $1/\gamma^2$ using the approximation that the total swept up mass is contained within the shell.

To simulate the encounter with a change in external medium, we model the external density profile as a piecewise function:
\begin{equation}
\rho_{ext} = \left\{\begin{array}{rl}
\rho_{wind}&\mbox{ $\gamma>\gamma_{enc}$} \\
\rho_{ISM}&\mbox{ $\gamma\leq\gamma_{enc}$}\end{array}\right.
\end{equation}
where $\gamma_{enc}$ is the fluid Lorentz factor at which the encounter with the new environment occurs. The dynamics before the encounter are described by the equations above, but the dynamics during the encounter and after the encounter are much different. The simulation does not necessarily return to a BM self similar solution for a relativistic blast wave in a homogeneous medium, for the BM solution is a slow attractor and a perturbation in the simulation, such as a density change, can cause large deviations from the BM solution \citep{Gruzinov2000}. Also, after the encounter, the blast wave may no longer be relativistic which prevents the blast wave from evolving toward the BM solution in a homogeneous medium, and a solution similar to ST is expected instead.

\subsection{Analytical Solution}\label{sec:AS}

To analytically model the dynamics of a blast wave in a stellar wind environment encountering a density change followed by a homogeneous medium, we extend the derivations done in PW and NG. With our extensions, the analytical solution is applicable for a wide range of density jumps as well as density drops.

\begin{figure}
\begin{center}
\includegraphics[width=0.49\textwidth]{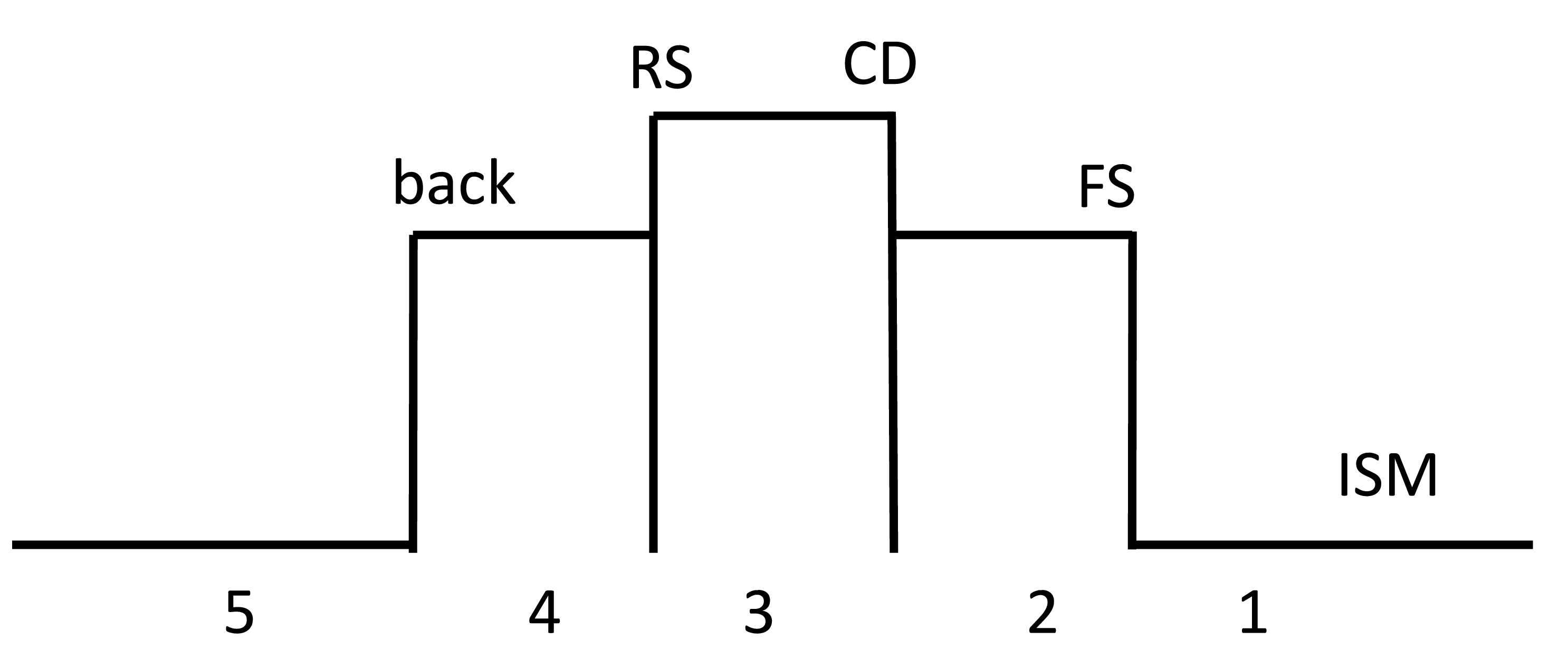}
\caption{Diagram of the density profile formed from a blast wave in a stellar wind encountering a ISM. The fluid in region 1 is the ISM into which the blast wave is traveling. Region 2 contains all of the mass swept up after the encounter with the density change. The fluid in regions 3 and 4 is the swept up mass prior to the encounter, meaning that after the encounter, the amount of mass contained in regions 3 and 4 is constant. The fluid in region 3 is that which has been shocked by the reverse shock, and the fluid in region 4 is the unshocked material that is unaware of the encounter.  \label{fig:shocks}}
\end{center}
\end{figure}

For simplicity, we model the shocks of the blast wave as simple homogeneous shells. During the encounter, the shock breaks up in to three different regions--regions 2, 3, and 4 of Figure~\ref{fig:shocks}.
To get the full analytical solution of the simulation, the pressures, densities, Lorentz factors, and radii of the three regions (regions 2, 3, and 4, which are referenced by subscripts from here forward in this paper) are needed.

Across the contact discontinuity, the pressure and fluid Lorentz factors are equal, meaning $p_2 = p_3$, and $\gamma_2 = \gamma_3 = \Gamma_{CD}$. Before the reverse shock passes the ``back'' of the shock,  the fluid in region 4 has no knowledge of the encounter, meaning the density, pressure, and Lorentz factor are calculated as if no encounter has occurred. The rest of the fluid has experienced the encounter, and to calculate the values analytically, a convenient factor to define is that of NG Equation 3, $\psi$, which can be derived from conservation of mass.
\begin{equation}
\psi^2 \equiv \left[\frac{\gamma_{4}(T_{enc})}{\gamma_2(T_{enc})}\right]^2 = \frac{3a - 4}{\sqrt{12/a}(a-1) - 1}
\end{equation}
\begin{equation}\label{eq:a}
a = \frac{\rho_{ISM}}{\rho_{wind}(R = R_{enc})} 
\end{equation}
Using this factor, the Lorentz factor of the fluid behind the forward shock is described by a modified version of Equation 6 from NG:
\begin{equation}\label{eq:g2}
\gamma_2 = \gamma_{enc}\left[\psi^2 + \left(\frac{17 - 4k}{17}  \right)a\left( \left(\frac{T}{T_{enc}}\right)^3 - 1\right) \right]^{-1/2}
\end{equation}
where Equation~\ref{eq:g2} differs from NG Equation 6 in that $\gamma_2 \propto \gamma_{enc}$ as opposed to $\gamma_2 \propto \gamma_{4}$ of NG Equation 6.

To calculate the radius of the forward shock to the leading order of $1/\gamma_2^2$, we integrate the velocity of the forward shock: 
 \begin{equation}\label{eq:RFS_int}
\begin{split}
R_{FS}(T) = &R_{enc} + c\int^T_{T_{enc}}\beta_{FS} dt \\
\cong  &R_{enc} + cT_{enc} a\frac{17 - 4k}{68\gamma_{enc}^2} \left( \frac{T}{T_{enc}} - \frac{3}{4} - \frac{T^4}{4T_{enc}^4}\right) \\
 &+ c T_{enc}\left(\frac{T}{T_{enc}} - 1\right)\left( 1 - \frac{\psi^2}{4\gamma_{enc}^2} \right).
\end{split}
\end{equation}
The radius of the contact discontinuity is calculated in the same fashion as Equation~\ref{eq:RFS_int}, except with $\beta_{FS}$ replaced by $\beta_{CD}$.

The pressure in regions 2 and 3 are identical, and are found using the strong shock jump conditions.
\begin{equation}
p_2 = p_3 = \frac{2}{3}\rho_{ISM}\Gamma_{FS}^2c^2
\end{equation}
One important aspect to note is that the reverse shock is not necessarily strong or relativistic in the frame of the fluid in region 4: if the forward shock is relativistic, the reverse shock is not (see PW for details). However, if the blast wave encounters a strong density jump causing the forward shock to become non-relativistic, the reverse shock can form with relativistic speeds. Since we cannot assume a relativistic reverse shock, we use the full shock jump conditions yielding the following equation for the density in region 3:
\begin{equation}
\rho_3 = \frac{4p_3\bar\gamma_3 \rho_4}{3p_4+ p_2},
\end{equation}
where $\bar\gamma_3$ is the Lorentz factor of the fluid in region 3 from the frame of the unshocked fluid in region 4:
\begin{equation}
\bar\gamma_3 = \frac{1}{2}\left[ \frac{\gamma_4}{\gamma_3} + \frac{\gamma_3}{\gamma_4}\right].
\end{equation}
Lastly, we need to calculate the velocity and radius of the reverse shock. We use conservation of mass to obtain the velocity of the reverse shock:
\begin{equation}\label{eq:Brs}
\beta_{RS} = \frac{\rho_3\gamma_3\beta_3 - \rho_4\gamma_4\beta_4}{\rho_3\gamma_3 -\rho_4\gamma_4}.
\end{equation}

To calculate the radius of the reverse shock, we use conservation of mass as well. The total mass in regions 3 and 4 is constant after the encounter, meaning the integral over these regions equals the swept up mass prior to the encounter.
\begin{equation}\label{eq:Rrs_int}
\int_{R_{back}}^{R_{RS}}\!r^2\rho^{\prime}_4\, \mathrm{d}r + \int_{R_{RS}}^{R_{CD}}\!r^2\rho^{\prime}_3\, \mathrm{d}r = \int_0^{R_{enc}}\! r^2 \rho_{wind}(r)\, \mathrm{d}r
\end{equation}
Equation~\ref{eq:Rrs_int} is important because we know the densities in regions 3 and 4, we have calculated $R_{back}$ from Equation~\ref{eq:rback}, we know $R_{CD}$ from Equation~\ref{eq:RFS_int} but with $\beta_{FS}$ replaced by $\beta_{CD}$, and we know the total mass swept up prior to the encounter (the right hand side of Equation~\ref{eq:Rrs_int}). This leaves us with an equation for the radius of the reverse shock:
\begin{equation}\label{eq:Rrs}
R_{RS}^3 (\rho^{\prime}_4 - \rho^{\prime}_3) = \frac{3}{3-k}R_{enc}^3\rho_{wind}(R_{enc}) + R_{back}^3\rho^{\prime}_4 - R_{CD}^3\rho^{\prime}_3.
\end{equation}
Equations~\ref{eq:Rrs_int} and~\ref{eq:Rrs} contain $\rho^{\prime}$, which is the density of the fluid in the lab frame, $\rho^{\prime} = \gamma\rho$, whereas $\rho$, without a prime, is the co-moving density. 

Solving for $R_{RS}$ in Equation~\ref{eq:Rrs} is only applicable when the reverse shock has not yet passed the back of the shock. Once the reverse shock passes the back of the shock, Equation~\ref{eq:Rrs_int} no longer applies, and the new integral to solve is:
\begin{equation}\label{eq:Rb_int}
\begin{split}
\int_{R_{back}}^{R_{CD}}\!r^2\rho^{\prime}_3\, \mathrm{d}r  &= \int_0^{R_{enc}}\! r^2 \rho_{wind}(r)\, \mathrm{d}r \\
R_{back}^3&= R_{CD}^3 - \frac{3}{3-k}R_{enc}^3\frac{\rho_{wind}(R_{enc})}{\rho^{\prime}_3}
\end{split}
\end{equation}
where we now solve for $R_{back}$ instead of $R_{RS}$, and $R_{back}$ is no longer calculated from Equation~\ref{eq:rback} because region 4 has been shocked by the reverse shock and has knowledge of the encounter. The blast wave continues on with these two regions of Equation~\ref{eq:Rb_int}--one containing the newly swept up mass in region 2, and one containing the mass swept up prior to the encounter--resulting in the blast wave never completely returning to the BM solution of a blast wave in an ISM.

In the following section, Section~\ref{sec:NR}, we compare our analytical solution with the numerical results for various types of encounters.

\subsection{Numerical Results}\label{sec:NR}

Our numerical simulations demonstrate the resulting dynamics of an afterglow blast wave in a stellar wind environment encountering a sudden change in density and an ISM environment. We use the relativistic adaptive mesh (\textsc{ram}) parallel RHD code \citep{Zhang2006, Zhang2009} with a second-order weighted scheme. We simulate two different size jumps (``walls''), of factors 4 and 100, and one drop, of factor $\frac{1}{100}$. We use these specific initial conditions to numerically simulate a wide range of encounters thought to potentially cause light curve flares. However, as we explain in Section~\ref{sec:LC}, none of these scenarios result in flares on the time scale of $0.1$ to $10$ days. Before delving into the resulting light curves, we first explain the dynamics and numerical results.

We set up six simulations all of which have a starting time and radius equivalent to the fluid Lorentz factor behind the front of the shock, $\gamma=15$. This starting Lorentz factor was chosen to ensure $\gamma >\frac{1}{\theta_0}$, where $\theta_0=0.1$ rad (typical for afterglow jets, \citealt{Frail2001}) for our simulations and is the half opening angle of the jet.
\begin{itemize}
\item Simulation {\it no encounter}: \\ $\rho_{ext} = \rho_0\left(\frac{r}{R_{ref}}\right)^{-2}$
\item Simulation $\gamma5a4$:  \\ $\rho_{ext} = \left\{\begin{array}{rl}
\rho_0\left(\frac{r}{R_{ref}}\right)^{-2} &\mbox{ $\gamma>5$} \\
4\rho_0\left(\frac{R_{\gamma=5}}{R_{ref}}\right)^{-2} &\mbox{ $\gamma\leq5$}\end{array}\right.$
\item Simulation $\gamma5a100$: \\ $\rho_{ext} = \left\{\begin{array}{rl}
\rho_0\left(\frac{r}{R_{ref}}\right)^{-2} &\mbox{ $\gamma>5$} \\
100\rho_0\left(\frac{R_{\gamma=5}}{R_{ref}}\right)^{-2} &\mbox{ $\gamma\leq5$}\end{array}\right.$
\item Simulation $\gamma10a4$: \\ $\rho_{ext} = \left\{\begin{array}{rl}
\rho_0\left(\frac{r}{R_{ref}}\right)^{-2} &\mbox{ $\gamma>10$} \\
4\rho_0\left(\frac{R_{\gamma=10}}{R_{ref}}\right)^{-2} &\mbox{ $\gamma\leq10$}\end{array}\right.$
\item Simulation $\gamma10a100$: \\$\rho_{ext} = \left\{\begin{array}{rl}
\rho_0\left(\frac{r}{R_{ref}}\right)^{-2} &\mbox{ $\gamma>10$} \\
100\rho_0\left(\frac{R_{\gamma=10}}{R_{ref}}\right)^{-2} &\mbox{ $\gamma\leq10$}\end{array}\right.$
\item Simulation $\gamma5a\frac{1}{100}$: \\ $\rho_{ext} = \left\{\begin{array}{rl}
\rho_0\left(\frac{r}{R_{ref}}\right)^{-2} &\mbox{ $\gamma>5$} \\
0.01\rho_0\left(\frac{R_{\gamma=5}}{R_{ref}}\right)^{-2} &\mbox{ $\gamma\leq5$}\end{array}\right.$
\end{itemize}

We first ran all six simulations in one dimension, and then ran three simulations in two dimensions, choosing to run those which would be most informative to answering the questions outlined in the Introduction. 

\begin{figure}
\begin{center}
\includegraphics[width=0.49\textwidth]{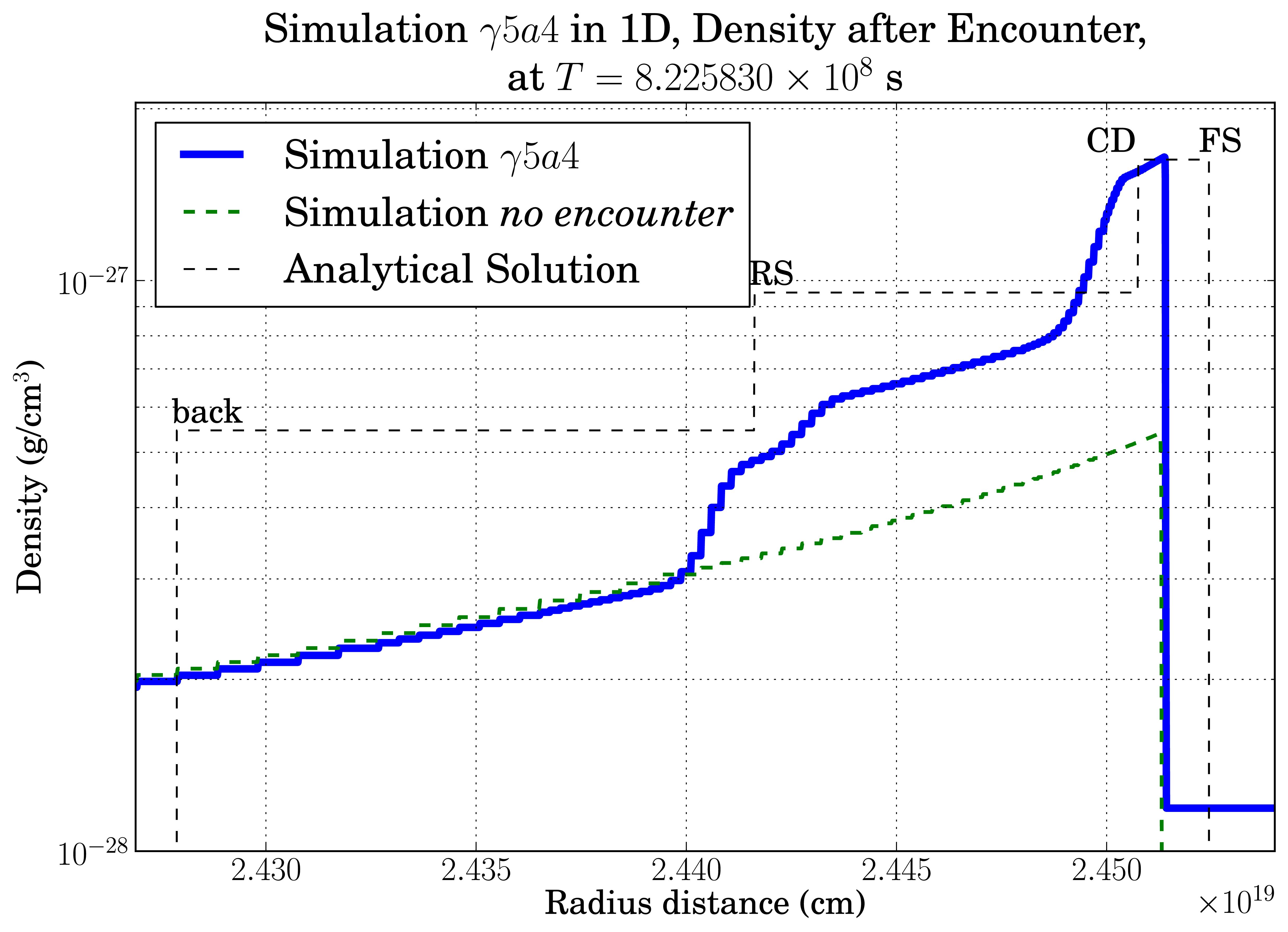}
\caption{Figure of the shocks formed during the encounter for simulation $\gamma5a4$ in 1D (wind termination shock at $\gamma_{enc} = 5$, and a density increase of factor 4). The simulation is plotted in blue, the dashed black line shows the analytical solution for this simulation, and the green dashed line represents simulation {\it no encounter} in 1D (no circumburst environment change) at the same time as simulation $\gamma5a4$ is depicted. \label{fig:dens_shocks_sim2}}
\end{center}
\end{figure}

Figure~\ref{fig:dens_shocks_sim2} shows the analytical solution of the density of simulation $\gamma5a4$, in one dimension plotted against the numerical result of simulation $\gamma5a4$ in one dimension, showing the correspondence of our analytical model with our numerical results of the shock formation at the encounter. Our analytical model assumes homogeneous shells with constant density between the shocks which is why the analytical solution in Figure~\ref{fig:dens_shocks_sim2} has a clearly marked ``back'' of the shock, whereas the numerical simulation uses the full BM profile and has no exact ``back'' of the shock. The small discrepancy between the analytical and numerical $R_{FS}$ shown in Figure~\ref{fig:dens_shocks_sim2} is roughly $\Delta{R_{FS}} = 9.6\times10^{15}$ cm. Our analytical solution is accurate to the order of $1/\gamma^2 = 0.0855$ at the time shown in Figure~\ref{fig:dens_shocks_sim2} and $\frac{\Delta{R_{FS}}}{cT} = 3.893\times10^{-4}\ll1/\gamma^2$, meaning that the analytical model conforms to the numerical results to the required accuracy. Also, it is apparent from Figure~\ref{fig:dens_shocks_sim2} that the densities within each region are not constant, but actually have a positive slope. This is a known discrepancy between the assumed homogeneous shells and the full BM solution, and leads to a slight over-prediction of the flux during the encounter by the analytical solution \citep{vanEerten2009windtermination}.

\begin{figure}
\begin{center}
\includegraphics[width=0.49\textwidth]{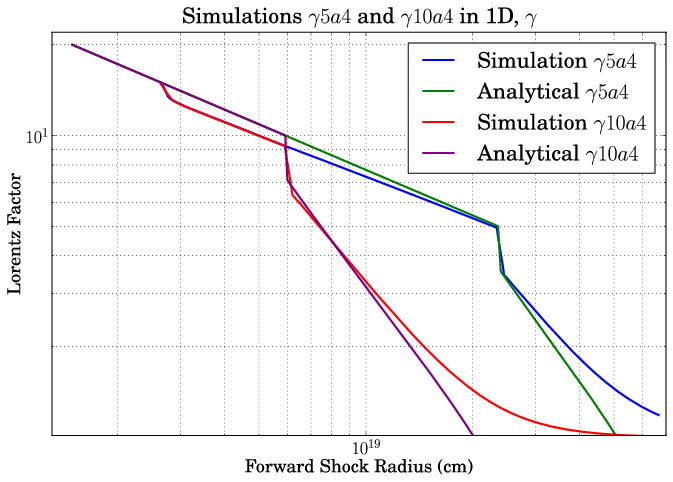}
\caption{Lorentz factor of the fluid behind the forward shock plotted over radius for the analytical solution as well as the numerical simulation for simulation $\gamma5a4$ (wind termination shock at $\gamma_{enc} = 5$, and a density increase of factor 4) and $\gamma10a4$ (wind termination shock at $\gamma_{enc} = 10$, and a density increase of factor 4) in 1D. \label{fig:lfac_v_r_sim2}}
\end{center}
\end{figure}

\begin{figure}
\begin{center}
\includegraphics[width=0.49\textwidth]{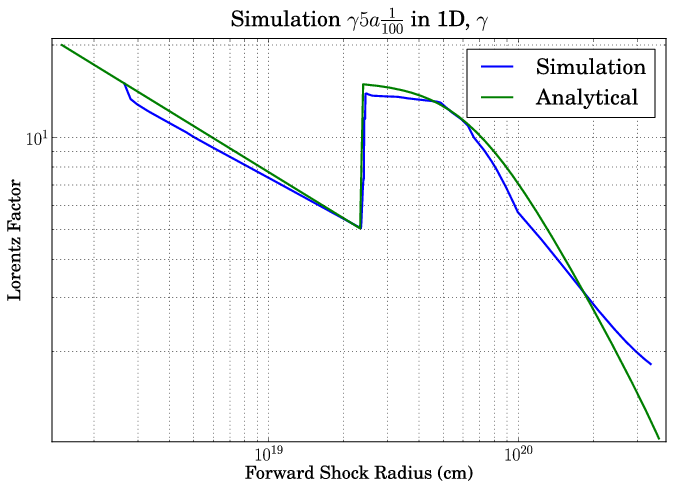}
\caption{Lorentz factor of the fluid behind the forward shock plotted over radius for the analytical solution and the numerical simulation for simulation $\gamma5a\frac{1}{100}$ in 1D (density decrease of factor 100 at $\gamma_{enc} = 5$).\label{fig:lfac_v_r_sim6}}
\end{center}
\end{figure}

Figures~\ref{fig:lfac_v_r_sim2} and~\ref{fig:lfac_v_r_sim6} show our analytical solutions plotted with the simulation results of the Lorentz factors behind the forward shocks as they evolve. The small deviations towards the beginning of the simulations (at smaller radii) in these figures are from lack of resolution. The resolution required to resolve the blast wave at these early times is very high. However, for the simulations with encounters at $\gamma_2 = 5$, the discrepancy disappears well before the encounter. For the simulations with the encounter at $\gamma_2 = 10$, the lack of resolution does affect the convergence of the numerical and analytical results, but qualitatively, the results are the same. The resolutions of our simulations are discussed further in Appendix~\ref{App:AppendixA}. The deviations at large radii (at very late time) in Figures~\ref{fig:lfac_v_r_sim2} and~\ref{fig:lfac_v_r_sim6} are from the Lorentz factor of the forward shock becoming non-relativistic, and our analytical model having accuracy of order $1/\gamma^2$.

From the examples shown in Figures~\ref{fig:dens_shocks_sim2} through~\ref{fig:lfac_v_r_sim6}, it is apparent that our analytical solution accurately predicts the dynamics of a one dimensional blast wave in a stellar wind encountering a change in density, be it a jump or drop, followed by an ISM environment. Also, it is important to note that when the post encounter forward shock remains relativistic, the fluid Lorentz factor at which the encounter occurs does not qualitatively change the post-encounter dynamics in one dimension, as seen in Figure~\ref{fig:lfac_v_r_sim2}. The post-encounter dynamics are simply scaled to a reflect the larger encounter Lorentz factor. This is also clear from the analytical model, since neither the change in Lorentz factor at the encounter nor the pre- and post-encounter slopes depend on Lorentz factor.

\begin{figure*}
\centering
\includegraphics[width=0.49\textwidth]{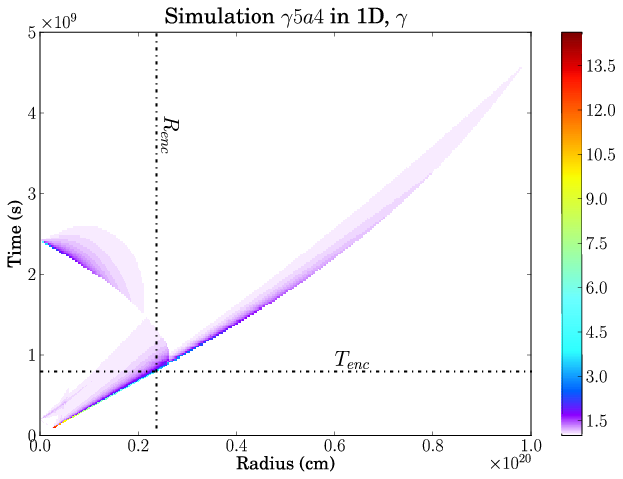}
\includegraphics[width=0.49\textwidth]{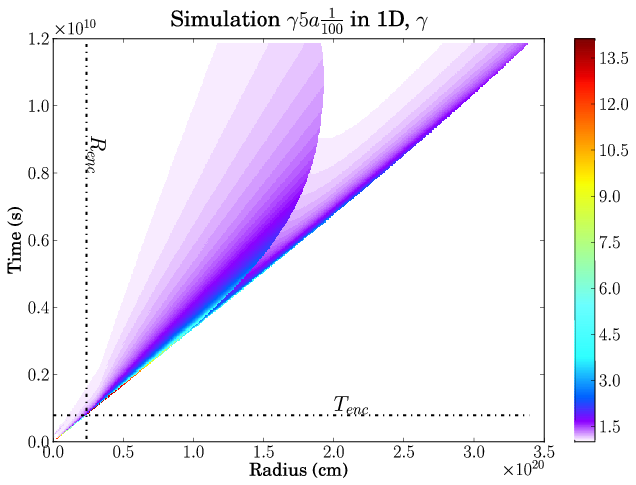}
\includegraphics[width=0.49\textwidth]{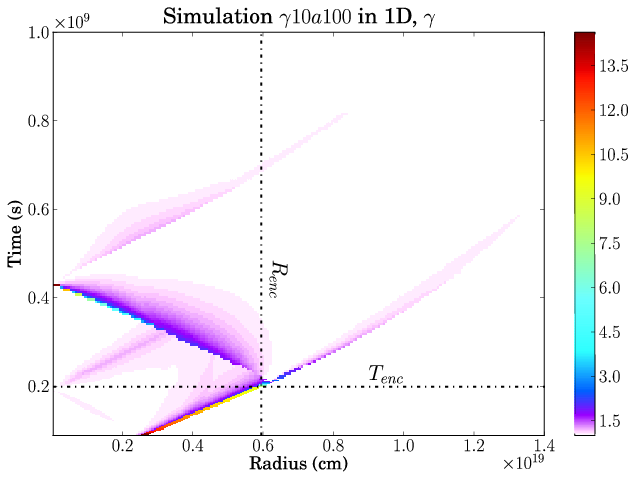}
\includegraphics[width=0.49\textwidth]{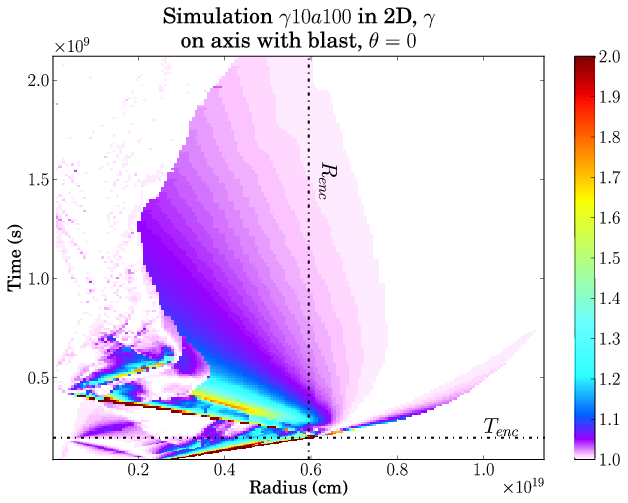}
\caption{Plot shows in color coding the fluid Lorentz factor, with radial distance on the horizontal axis and time on the vertical axis. For one dimensional simulations, the Lorentz factor does not depend on angle, but for the two dimensional simulation, only the Lorentz factor along the axis through the center of the blast wave is shown. Top left: simulation $\gamma5a4$ in 1D (blast wave in stellar wind that encounters a density jump of factor 4 at $\gamma_{enc}=5$). Top right: simulation $\gamma5a\frac{1}{100}$ in 1D (blast wave traveling in a stellar wind environment that encounters a density drop of factor 100 at $\gamma_{enc}=5$). Bottom: simulation $\gamma10a100$ in 1D (left) and an on axis (with the blast wave) slice of the 2D simulation (right, blast wave traveling in a stellar wind environment encountering a density jump of a factor of 100 at $\gamma_{enc}=10$). \label{fig:sim_gamma}}
\end{figure*}

The one dimensional simulations show that the dynamics follow the analytical predictions that the Lorentz factor of the forward shock drops immediately as it encounters a higher density region, and the amount by which it drops is proportional to the change in density at the encounter. This is shown in Figure~\ref{fig:sim_gamma} for simulations $\gamma5a4$,  $\gamma5a\frac{1}{100}$, and $\gamma10a100$. Figure~\ref{fig:sim_gamma} depicts the Lorentz factor in the color coding, with the complete radial fluid profile along the central axis of the blast wave (plotted in the horizontal) over time (plotted in the vertical). This figure shows the evolution of the forward shock over time before and after the encounter as well as the formation of the reverse shock at the encounter and its post-encounter evolution. Figure~\ref{fig:sim_gamma} clearly illustrates the varying behaviors of the forward and reverse shocks for the various simulations. Simulations $\gamma5a100$ and $\gamma10a4$ show similar behavior. The left column in Figure~\ref{fig:sim_gamma} shows the Lorentz factor over time and radius for simulations $\gamma5a4$ and $\gamma10a100$ in 1D.  If the blast wave encounters a drop in density, the Lorentz factor of the blast wave will increase (i.e., the blast wave will speed up), again, in proportion to the size of the drop, which is seen in the top right panel in Figure~\ref{fig:sim_gamma}.

The effects of the magnitude of density change on the dynamics of the reverse shock are shown in Figure~\ref{fig:sim_gamma}.  The top left panel in Figure~\ref{fig:sim_gamma} depicts the Lorentz factor for the simulation with a small density increase of factor 4. Shortly after the encounter, the reverse shock slowly travels away from the forward shock towards the origin, but this is not an immediate process like that of a larger density jump shown in the bottom left panel in Figure~\ref{fig:sim_gamma}. For a large density jump, almost immediately after the encounter, the reverse shock has a large enough Lorentz factor, compared to the Lorentz factor of the forward shock, to break away from the forward moving fluid and travel backwards towards the origin. This difference in reverse shock Lorentz factors of simulation $\gamma5a4$ and simulation $\gamma10a100$ is from the forward shock becoming immediately non-relativistic after the encounter with a large density jump (of factor 100) versus the much smaller decrease in forward shock Lorentz factor with a smaller density jump (of factor 4). As explained in Section~\ref{sec:AS}, the amount by which the forward shock Lorentz factor drops is proportional to the Lorentz factor with which the reverse shock forms. For simulation $\gamma5a\frac{1}{100}$, which has a density decrease of factor 100, the reverse shock takes much longer to overcome the forward moving fluid and start traveling backwards towards to origin. This is seen in the top right panel in Figure~\ref{fig:sim_gamma}. The Lorentz factor of the forward shock increases, causing the Lorentz factor of the reverse shock shortly after the encounter to be very small in comparison with the forward shock Lorentz factor.

A closer look at the bottom left panel in Figure~\ref{fig:sim_gamma} reveals the reflecting-like behavior of the inner boundary as the reverse shock travels towards the origin and is seemingly reflected back. This behavior is independent of boundary conditions or inner boundary position and results from the increase in pressure caused by the reverse shock traveling toward the origin. This increase in pressure creates conditions similar to a fireball, causing a new blast wave to be formed. This is not seen with simulations $\gamma5a4$, $\gamma10a4$, or $\gamma5a\frac{1}{100}$ which have smaller density jump factors/a density drop, but is seen with simulation $\gamma5a100$ (wind termination shock when $\gamma_2 = 5$ with a density jump of 100) yielding the conclusion that even at slower speeds, the strong, large density jump still slows the forward shock enough to cause a strong reverse shock to form and in turn, cause the conditions similar to a fireball. From here forward in this paper, we will refer to this newly formed blast wave caused by the strong reverse shock as the ``secondary blast wave'' for simplicity and clarity. 

The formation of the secondary blast wave is also seen in our two dimensional model of simulation $\gamma10a100$.  However, the fluid instabilities that follow the encounter with the wind termination shock in two dimensions slow the secondary blast wave causing it to never encounter the wind termination shock, and in turn, the continuous reflecting back and forth is never realized. This is shown in the bottom right panel in Figure~\ref{fig:sim_gamma}, which displays the fluid Lorentz factor over time and radius of a one dimensional slice of the two dimensional simulation $\gamma10a100$ along the axis of the jet.

\begin{figure*}
\centering
\includegraphics[width=0.49\textwidth]{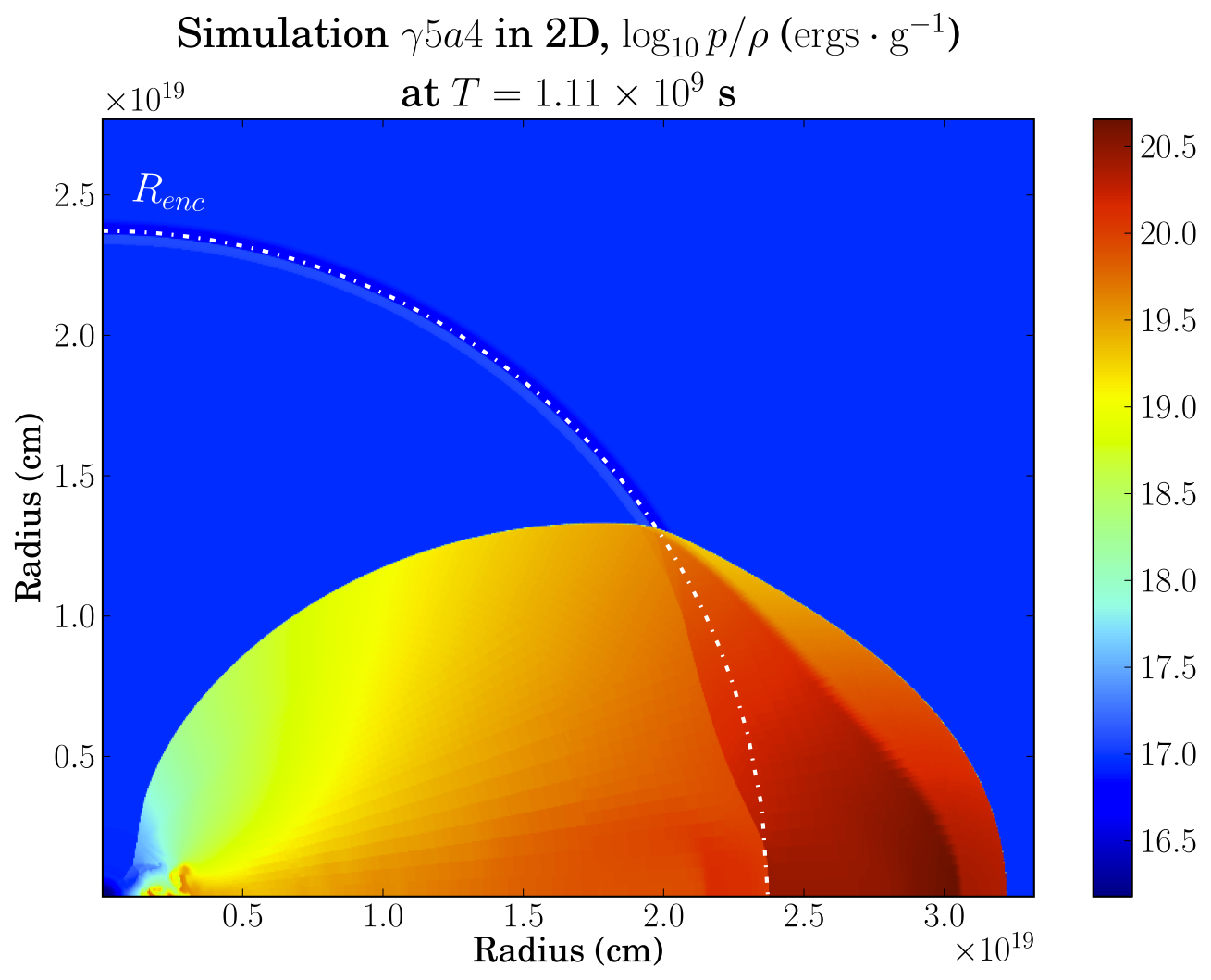}
\includegraphics[width=0.49\textwidth]{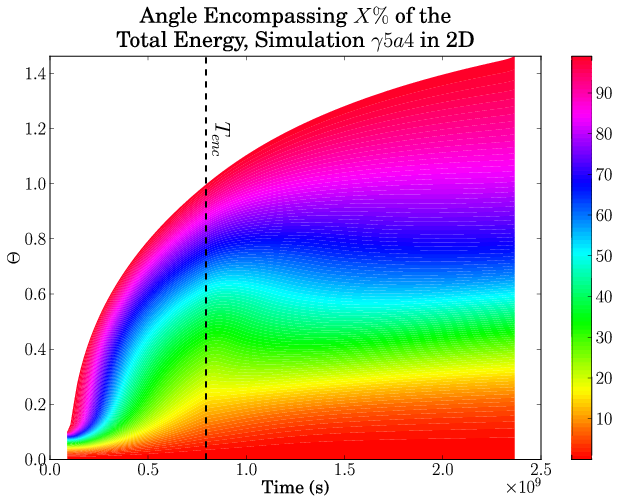}
\includegraphics[width=0.49\textwidth]{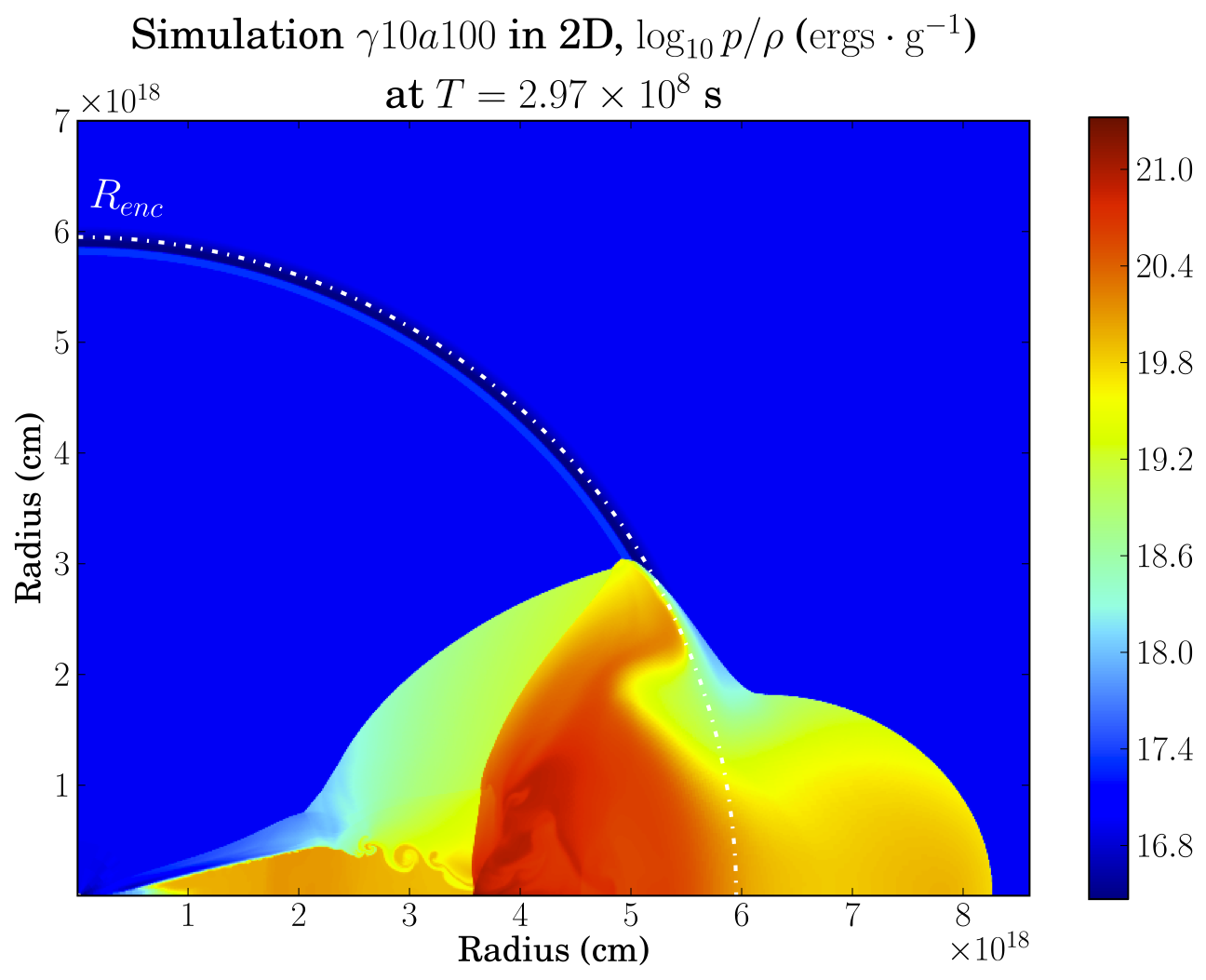}
\includegraphics[width=0.49\textwidth]{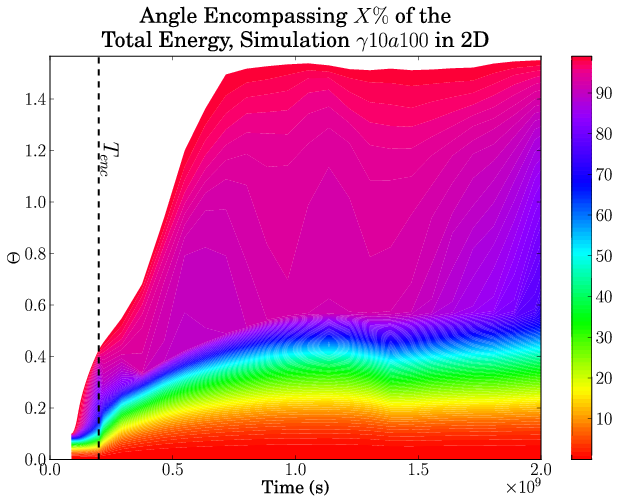}
\includegraphics[width=0.49\textwidth]{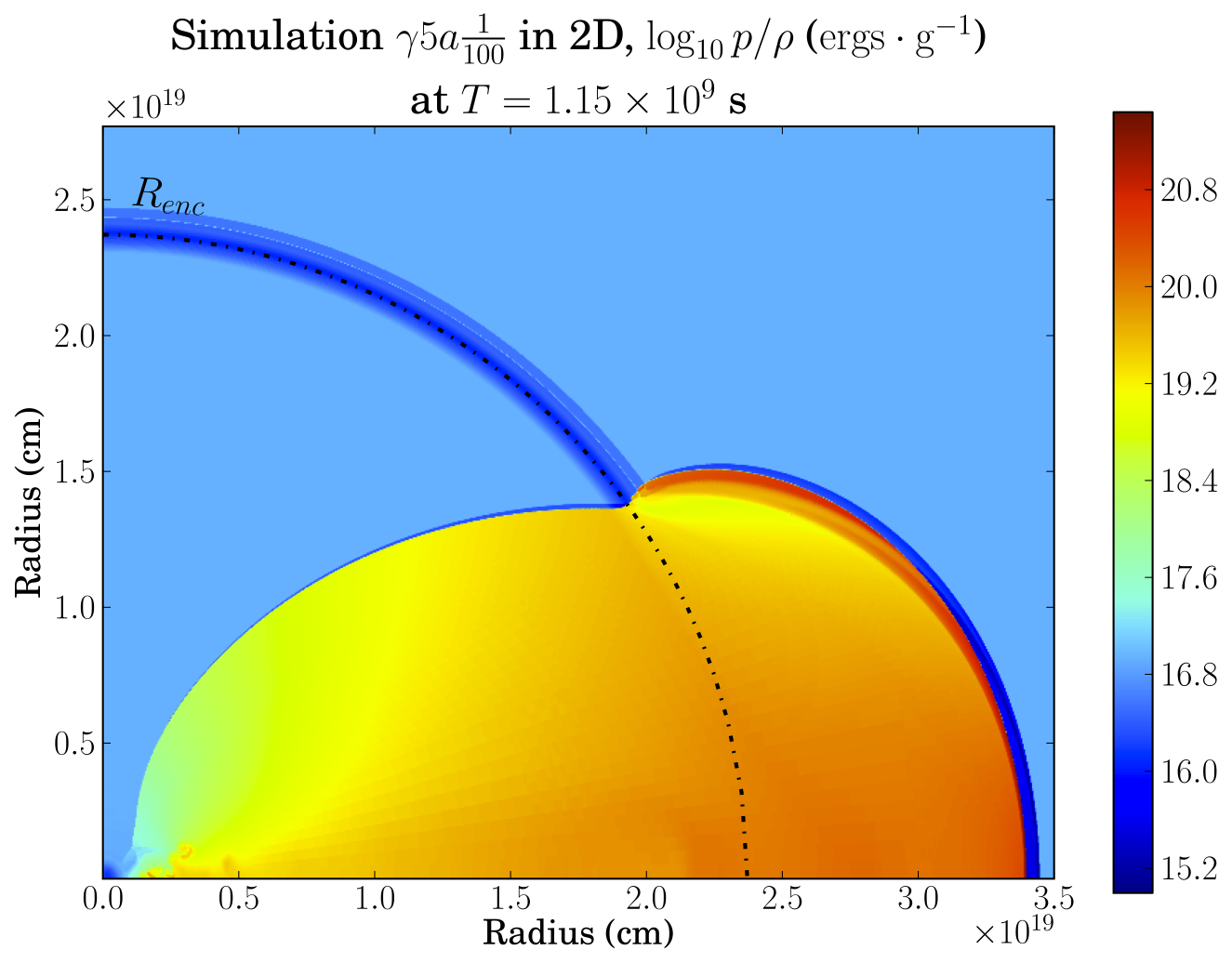}
\includegraphics[width=0.49\textwidth]{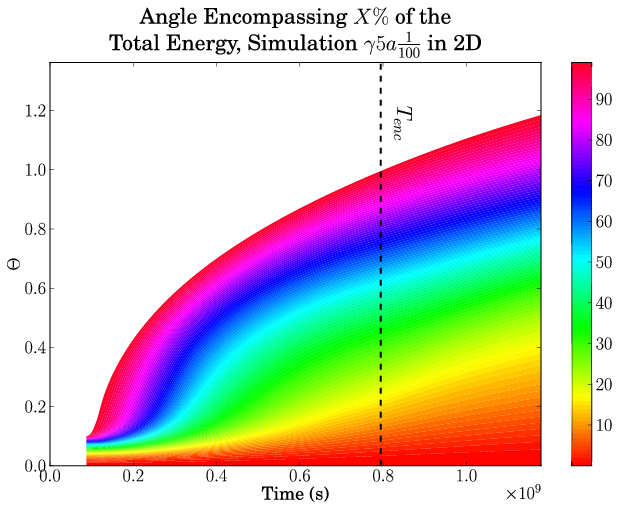}
\caption{Figures of the two dimensional dynamics. Left column: plots of the logarithm of specific energy at a time after the encounter. The blast wave is aligned with the horizontal axis. Right column: figures showing the angle enclosing various percentages of the blast wave energy over time. The horizontal line denotes the time of the encounter. Top row: simulation $\gamma5a4$ in 2D (blast wave traveling in a stellar wind environment encountering a wind termination shock at $\gamma_{enc}= 5$ and a density increase of factor 4). Middle row: simulation $\gamma10a100$ in 2D (blast wave traveling in a stellar wind environment encountering a wind termination shock at $\gamma_{enc}= 10$ and a density increase of factor 100). Bottom row: simulation $\gamma5a\frac{1}{100}$ in 2D (blast wave traveling in a stellar wind environment encountering a wind termination shock at $\gamma_{enc}= 5$ and a density decrease of factor 100). \label{fig:sim_energy}}
\end{figure*}

In order to study the effect of the density change on the post encounter blast wave collimation, we ran simulations $\gamma5a4$, $\gamma10a100$, and $\gamma5a\frac{1}{100}$ in two dimensions. Figure~\ref{fig:sim_energy} shows the results of our two dimensional simulations. We find that when the blast wave encounters a sudden density increase that is not very large (of factor 4, like that of simulations $\gamma5a4$ and $\gamma10a4$), the fastest, most energetic fluid will puncture the wall and continue through to the new medium (top left panel in Figure~\ref{fig:sim_energy}). This behavior results in the blast wave recollimating after the encounter as only the fluid with the highest pressure can pass through to the new region and open a pathway for the rest of the blast wave to continue through to the new environment. This is most clearly depicted in the top right panel in Figure~\ref{fig:sim_energy}, which shows the percentage of energy enclosed at various angles over time for simulation $\gamma5a4$ in two dimensions. Before the encounter, the blast wave spreads out, which is seen as the angles encompassing certain percentages of energy increase over time at times less then $T_{enc}$ in the top right panel in Figure~\ref{fig:sim_energy}. At the time of the encounter, the most energetic fluid passes through to the new environment and recollimates, which compresses a large percentage of the blast wave energy into a smaller angle. This accounts for the decrease in angle encompassing roughly half of the total energy after the encounter in the top right panel in Figure~\ref{fig:sim_energy}. Also, if there were no encounter, the angle encompassing roughly $80\%$ of the total energy would continue to increase as it did prior to the encounter in the top right panel in Figure~\ref{fig:sim_energy}, but instead, the angle stays relatively constant. This is because the fluid of higher energy (depicted with colors of yellow through red in the top left panel in Figure~\ref{fig:sim_energy}) no longer spreads sideways but tries to travel through the wall to the new medium. The only part of the blast wave that continues to spread sideways is the lowest energy fluid (shown in green and light blue in the top left panel in Figure~\ref{fig:sim_energy}), which results in the continual increase of the angle encompassing $100\%$ of the total energy of the blast wave for simulation $\gamma5a4$ in 2D (shown in  the top right panel in Figure~\ref{fig:sim_energy}) after the encounter.

\begin{figure}
\begin{center}
\includegraphics[width=0.49\textwidth]{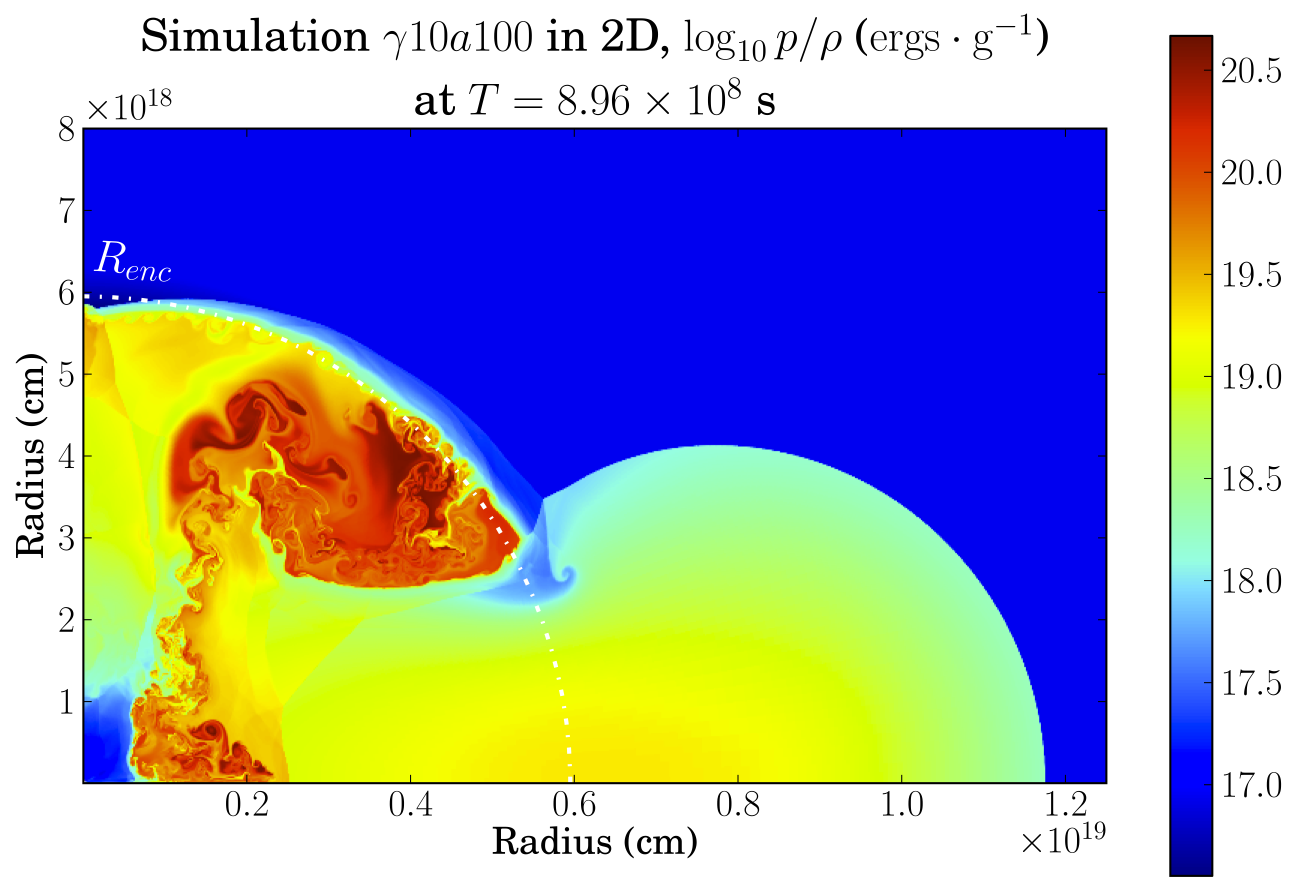}
\caption{Figure of the specific energy of the 2D simulation $\gamma10a100$ (blast wave traveling in a stellar wind environment encountering a wind termination shock at $\gamma_{enc} = 10$ and a density increase of factor 100) shown at a later time than the middle left panel in Figure~\ref{fig:sim_energy}. The blast wave in this figure is aligned with the horizontal axis.\label{fig:sim5_2d_se_2}}
\end{center}
\end{figure}

The dynamics of a blast wave encountering a large density increase (on the order of 100, like that of simulations $\gamma5a100$ and $\gamma10a100$) are different than that of a smaller density increase. The most energetic fluid punctures the wall and continues on to the next medium, similar to the most energetic fluid encountering a smaller density increase, but instead of opening up a pathway for the rest of the blast wave to move through, a strong reverse shock is formed. This strong reverse shock shocks and spreads the fluid that has not yet passed through to the new medium, forming vortices that are trapped within the stellar wind environment and never pass through to the ISM. The middle left panel in Figure~\ref{fig:sim_energy} and Figure~\ref{fig:sim5_2d_se_2} illustrate these dynamics showing the specific energy of simulation $\gamma10a100$ in two dimensions at two times after the encounter. The middle left panel in Figure~\ref{fig:sim_energy} depicts the blast wave a short time after the encounter showing that the most energetic fluid has passed through to the ISM, and the reverse shock is traveling back through the fluid in the stellar wind, shocking and spreading the fluid. At the much later time of Figure~\ref{fig:sim5_2d_se_2}, the highest energy fluid that passed through to the ISM has began to expand as it assimilates to the new medium. Also, the reverse shock has now passed through the fluid behind the density jump, caused the formation of vortices, and created the condition for the secondary blast wave to form. The remains of secondary blast wave can be seen in Figure~\ref{fig:sim5_2d_se_2} by the vortices near the horizontal axis at radii around $2\times10^{18}$ cm.

The middle right panel in Figure~\ref{fig:sim_energy} shows the percentages of energy encompassed in various angles, depicting that the highest concentration of energy is contained within the angle holding the fluid that passes through the density jump, and a very minimal amount of energy is contained within the vortices that have been formed in the stellar wind region. The chaotic behavior of the region in purple and pink in the middle right panel in Figure~\ref{fig:sim_energy} is from those vortices as well.

Lastly, the dynamics of the simulation with the sudden density drop do not show a recollimation with the increase in Lorentz factor. The fluid speeds up and as it does, it spreads at a rate slightly faster than in the stellar wind environment. This is shown in the bottom row of Figure~\ref{fig:sim_energy}.

With the understanding of the dynamics of the blast wave encountering a sudden change in density, the next section discusses the resulting light curves.

\section{Light Curves}\label{sec:LC}

We calculate the GRB afterglow light curves using the radiation calculation methodology from~\cite{vanEerten2010offaxis, vanEerten2010transrelativistic} at an X-ray frequency of $5\times10^{17}$ Hz (except where otherwise specified), similar to the frequency detected by {\it Swift}. The time at which the flux from the local emission of the afterglow is observed is denoted as the ``observer time,'' or $t_{obs}$, and is calculated by:
\begin{equation}
t_{obs} = t_{e} - \frac{r\cos{\theta}}{c}
\end{equation}
where $t_e$ is the emission time (i.e., the lab frame time of the emitting fluid element), $r$ is the radius of the local fluid element at $t_e$, and $\theta$ is the angle between the direction to the fluid element and the direction to the observer. For an emission time, $t_e$, the fluid at the front of the shock will be observed first, and can be represented as:
\begin{equation}\label{eq:tobs1}
T_{obs}\cong t_{e} - \frac{R}{c}
\end{equation}
where $T_{obs}$ denotes the earliest time any flux from the blast wave at the time, $t_e$, is observed. On our figures, we label the first time the encounter can be observed as $T_{obs,enc}$.

For our light curves, we use values of $\epsilon_B = 0.01$, $\epsilon_e = 0.1$, and $p = 2.5$, where $\epsilon_B$ and $\epsilon_e$ are the fractions of internal energy that contribute to the magnetic field at the shock front and to accelerating electrons respectively, and $p$ expresses the energy distribution index of the shock-accelerated particles (and is not to be confused with pressure). We also do not consider electron cooling in our light curves shown here, but have found that this does not qualitatively change our results. We expect the pre-jet break light curves to follow the temporal behavior of $t^{(1-3p)/4}$ in the stellar wind environment, and $t^{3(1-p)/4}$ in the ISM (e.g., \citealt{Granot2002}).

It has been claimed that blast waves encountering wind termination shocks generate flares or rebrightenings which can be seen at early times and are caused from either the reverse shock forming at the encounter (PW) or the blast wave transitioning to the new medium \citep{Eldridge2006, Mesler2012}. As discussed below, from our analytical and numerical analysis,we find that a GRB jet encountering a change in circumburst medium does not cause a flare of the kind seen by {\it Swift}. Neither the reverse shock at the encounter, nor the blast wave's transition to the new medium at the encounter will cause an observable flare.  

\subsection{No Flares during the Observed Time Scale}

\begin{figure*}
\centering
\includegraphics[width=0.49\textwidth]{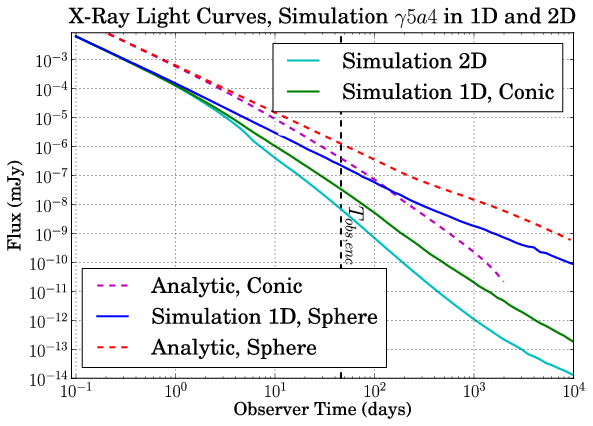}
\includegraphics[width=0.49\textwidth]{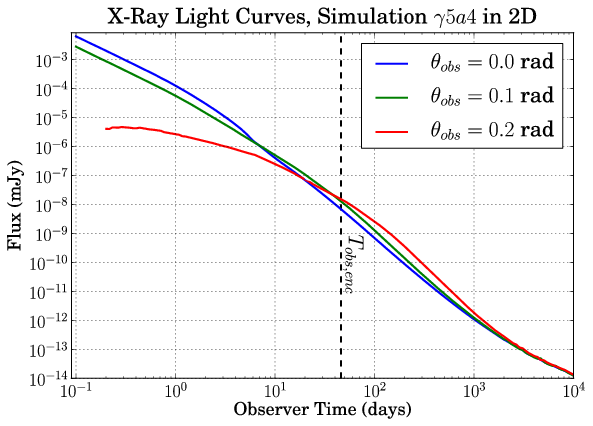}
\includegraphics[width=0.49\textwidth]{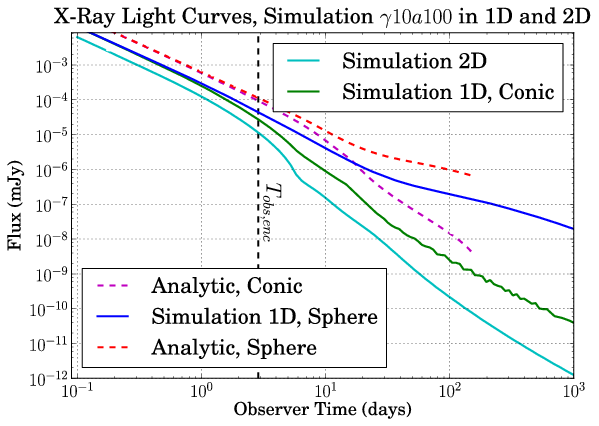}
\includegraphics[width=0.49\textwidth]{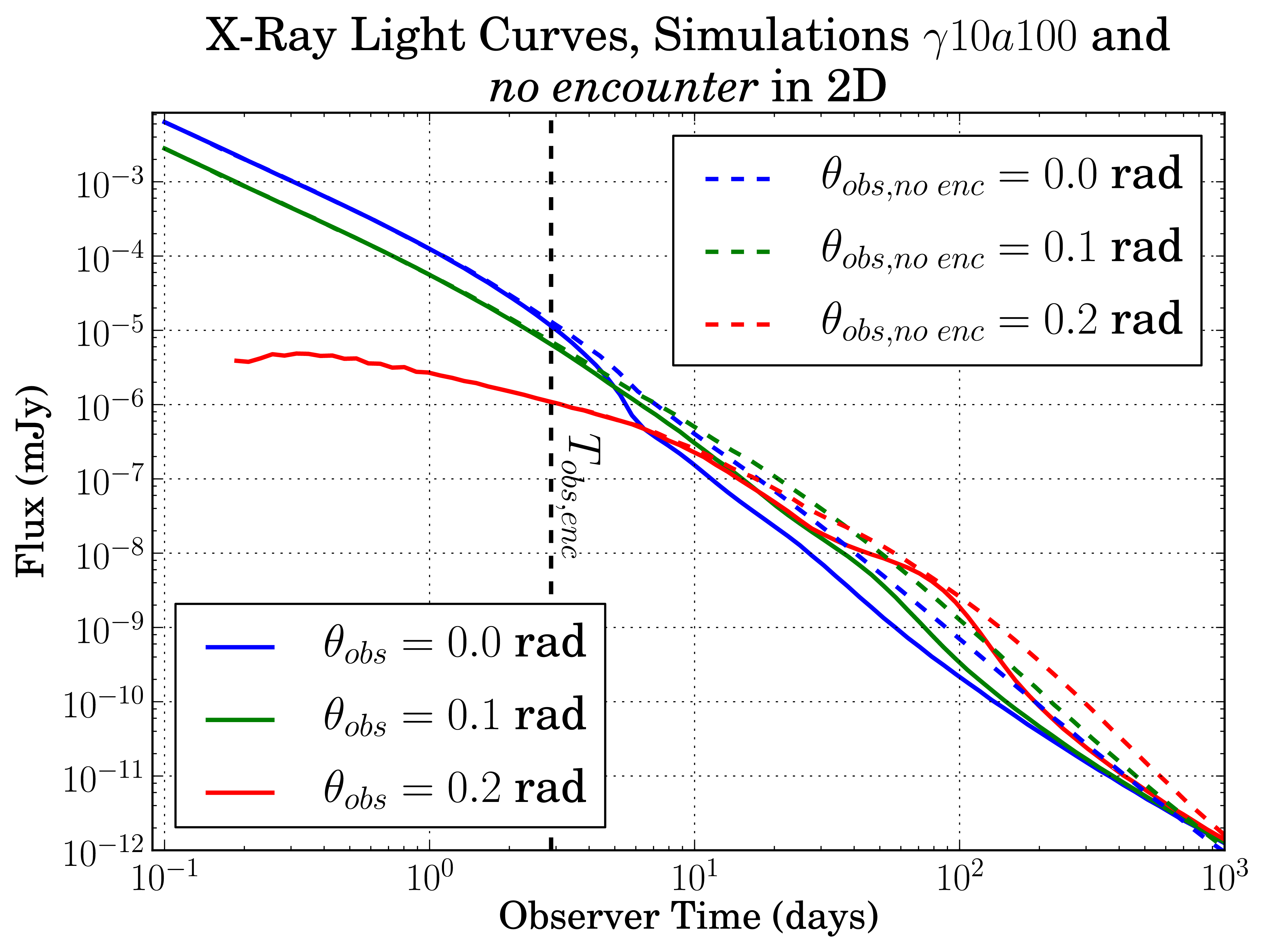}
\includegraphics[width=0.49\textwidth]{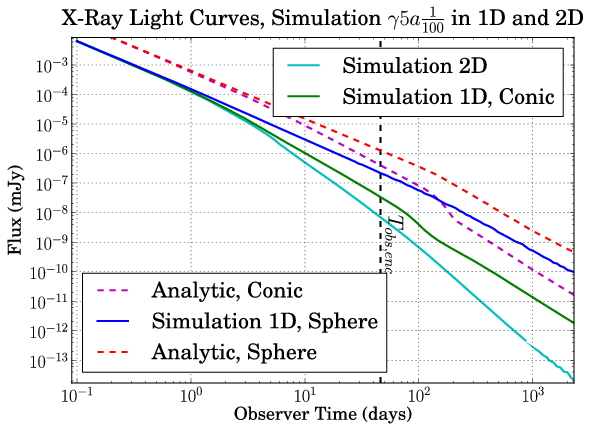}
\includegraphics[width=0.49\textwidth]{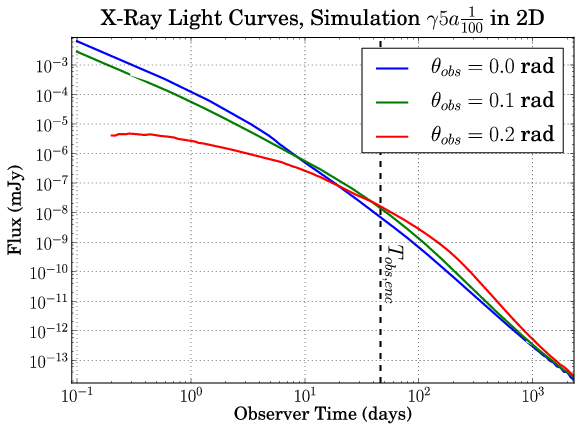}
\caption{Figures showing light curves calculated from simulations all with the jet half-opening angle of $\theta_0 = 0.1$. Left column: light curves for the simulations in one and two dimensions as well as the light curves calculated from the analytical model in one dimension. The conic and spherical one dimensional dynamics represent the assumption of assuming either spherical or conic outflow, respectively, when calculating the light curve. Right column: light curves calculated from the two dimensional simulations for various observer angles. Top row: light curves for simulation $\gamma5a4$ (blast wave encountering a density jump of factor 4 at an observer time of roughly 46 days). Middle row: light curves for simulation $\gamma10a100$ (blast wave encountering a density jump of factor 100 at an observer time of roughly 3 days). Right plot also shows the light curves for simulation {\it no encounter} at various observer angles. Bottom row: light curves for simulation $\gamma5a\frac{1}{100}$ (blast wave encountering a density drop of factor 100 at an observer time of roughly 46 days)\label{fig:sim_lc}}
\end{figure*}

Figure~\ref{fig:sim_lc} depicts that there are no observable flares in the flux emitted from a blast wave encountering a small jump, large jump, or large drop. This figure shows the light curves calculated from the one dimensional simulation and analytical model as well as the light curves calculated from the two dimensional numerical simulations for on and off-axis observer angles. For all cases studied in this paper, there were no observable flares.

\begin{figure*}
\includegraphics[width=0.49\textwidth]{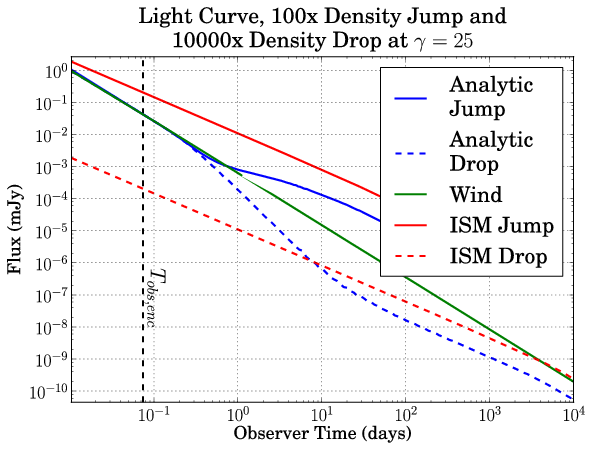}
\includegraphics[width=0.49\textwidth]{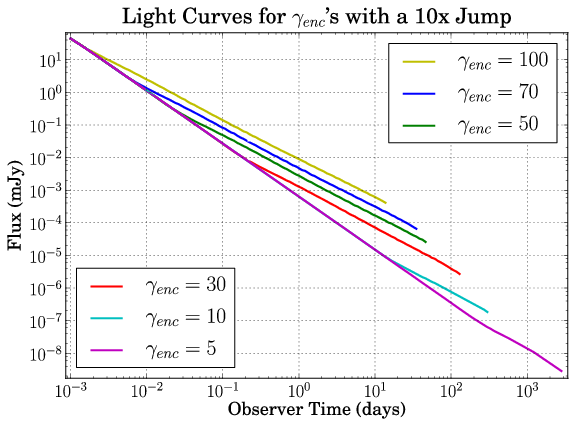}
\includegraphics[width=0.49\textwidth]{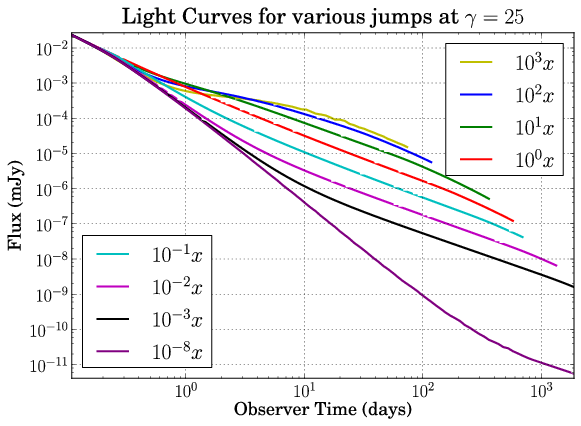}
\hspace{7 pt}\includegraphics[width=0.49\textwidth]{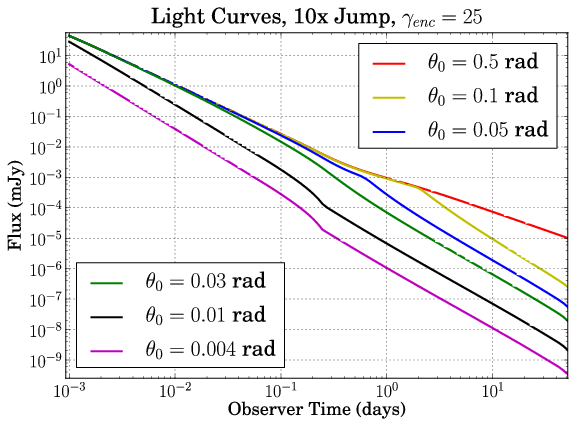}
\caption{Figures showing the transitions of light curves as the blast wave travels through an encounter with a density change. All figures are of light curves calculated from the one dimensional analytical model, and all except for the bottom right assume spherical outflow. Top left: blue light curves are of blast wave traveling in a stellar wind environment encountering a density jump of factor 100 (solid blue line) and a density drop of factor 10000 (dashed blue line) followed by a homogeneous medium. These are plotted against the light curve for a blast wave traveling solely in a stellar wind environment (green line) and the light curves for blast waves traveling solely in the homogeneous medium of the jump (red solid line) and of the drop (dashed solid line). Top right: light curves for blast waves traveling in a stellar wind encountering a density jump of factor 10 followed by a homogeneous medium with the encounters at various times. Bottom left: light curves for blast waves traveling in a stellar wind environment encountering a density change of various magnitudes at a time corresponding to $\gamma_{enc}=25$. Bottom right: light curves for blast waves traveling in a stellar wind environment encountering a density jump of factor 10 when $\gamma_{enc}=25$ for various opening angles. \label{fig:scaling}}
\end{figure*}

The simulations demonstrate that a blast wave encountering a wind termination shock of a small increase in density (of factor 4) does not result in any disruption in the flux observed at early times. In fact, for a jump of factor 4 (simulations $\gamma5a4$ and $\gamma10a4$), the light curve gradually transitions to the slope expected in the new environment, confirming the results from the spherical cases of NG and \cite{vanEerten2009windtermination}. This is seen with the on-axis light curves of the one and two dimensional simulations (top left panel in Figure~\ref{fig:sim_lc}) and with the light curves for a wide range of observer angles of the two dimensional simulation $\gamma5a4$ (top right panel in Figure~\ref{fig:sim_lc}). The encounter is first observable at roughly 46 days for the simulations with $\gamma_{enc} = 5$, and for our simulations of $\gamma_{enc}=10$, the first time the encounter is observed,  $T_{obs,enc}\approx3$ days. These times are much later than any flares observed from {\it Swift} and follow from our choice for $\gamma_{enc}$, which was selected for numerical reasons. However, the time at which the encounter is observed is inversely proportional to $\Gamma_{enc}^4$:
\begin{equation}
T_{obs, enc} = \frac{T_0 \Gamma_0^2}{4\Gamma_{enc}^4}.
\end{equation}
This scaling is calculated using Equations~\ref{eq:renc} and~\ref{eq:gamma_enc}, and noting that at (and before) the encounter $k=2$. It follows that modeling encounters at larger Lorentz factors, of say $\gamma_{enc}=25$, which corresponds to $T_{obs,enc}\approx0.07$ days, would result in the observed encounter being in the range of times flares are observed (see e.g., \citealt{Zhang2006b}). Nevertheless, from the top right panel in Figure~\ref{fig:scaling}, it is apparent that the light curves for blast waves encountering changes in external density profiles at varying Lorentz factors scale for one dimensional dynamics. Thus the light curve behavior at the time of the encounter for our numerical simulations of $\gamma_{enc}=5\;\;\mathrm{and}\;\;10$ are indicative of the behavior of encounters at larger Lorentz factors. To ensure that this is accurate, we analytically model encounters at $\gamma_{enc}=25$, and discuss the light curves from those dynamics throughout this section. The light curves calculated from our analytical model use the same linear radiative transfer algorithm as the light curves calculated from the numerical simulations.

For simulation $\gamma5a\frac{1}{100}$, the flux observed follows the same qualitative results as simulation $\gamma5a4$ at early times--the observed flux transitions smoothly to the slope of the flux observed solely in the ISM environment after the encounter. Even though the blast wave's velocity increases after the encounter (Figures~\ref{fig:lfac_v_r_sim6} and top right in~\ref{fig:sim_gamma}), a flare during the encounter is not observed. Instead, the flux drops more steeply as it evolves towards its lower post-encounter base level. This is shown in the two figures in the bottom row in Figure~\ref{fig:sim_lc} where the figure on the bottom left depicts the flux observed on axis with the blast wave using the one and two dimensional simulations along with the analytical solution, and the latter displays the flux observed at various observer angles for the two dimensional simulation.

It is apparent from the light curves of simulations $\gamma5a4$ and $\gamma5a\frac{1}{100}$, that as the magnitude of the flux of a blast wave in the ISM environment lowers, there is no rebrightening. Next, we discuss the light curves of simulation $\gamma10a100$, which are of a blast wave encountering an ISM environment of much higher density. 

With the magnitude of the flux of a blast wave traveling solely in the ISM environment being much higher than the magnitude of the flux from a blast wave in the wind (because the magnitude is a function of the density in that region), one might first assume a flare must be observed as the flux from the blast wave at the encounter evolves to the new slope and magnitude. However, this is not correct--there is {\it still no rebrightening} at the time of the encounter. This is shown in the middle row of Figure~\ref{fig:sim_lc}. The right middle panel in Figure~\ref{fig:sim_lc} shows the light curves for the 2D simulation $\gamma10a100$ plotted against the light curves for the 2D simulation {\it no encounter} for various observer angles. This figure illustrates the deviation of the light curve from the blast wave encountering the density jump from the light curve of no encounter at all. The magnitude of the flux decreases as the blast wave travels into the new medium resulting in no flaring activity. The off axis light curve calculated at $\theta_{obs}=0.2$ radians in the right panel in the middle row of Figure~\ref{fig:sim_lc} shows a bump at around 100 days. Although the flux does not increase at this point, it does deviate from the expected slope. This bump in the light curve is observable at a wide range of frequencies, from the optical to the X-ray, and exhibits similar behavior at all frequencies at which it is observed. The bump could be considered a rebrightening, however, it is only seen at observer angles larger than the jet opening angle, and it is unlikely that the prompt emission will be observed at this angle due to the extreme beaming needed of the gamma radiation.

Figure~\ref{fig:scaling} depicts the transition of the light curve as the blast wave changes environments as well as the overall scaling of the light curves. The top left panel in Figure~\ref{fig:scaling} shows analytical light curves for a blast wave traveling in a stellar wind environment with $\gamma_{enc}=25$ for two scenarios: a density jump of factor 100 (the solid blue line) and a density drop of factor 10000 (the dashed blue line). These light curves are plotted against the light curve for a blast wave traveling solely in the stellar wind environment (green line), a blast wave traveling solely in the homogeneous medium corresponding to the density jump of factor 100 (red solid line), and a blast wave traveling solely in the homogeneous medium corresponding to the density drop of factor 10000 (dashed red line). As seen in this figure, the observed flux after the encounter slowly transitions to the slope of the flux observed from the ISM environment, resulting in no sudden increase in flux. The flux does not immediately jump up to the new magnitude because the blast wave has been slowed by the encounter and is not energetic enough to cause a rebrightening. For the case of a drop--the flux observed from a blast wave in the new ISM environment is at a lower magnitude than the flux observed from a blast wave in the wind environment. This results in the flux observed from the blast wave encountering the new environment to decrease, instead of increase, after the encounter to evolve in the new environment.

In addition, this figure shows an interesting feature, that can also be seen in Figure~\ref{fig:sim_lc}, of a shallowing of the light curve. The light curve transitions from the slope of a blast wave in a stellar wind to the less steep slope of a blast wave in an ISM. This late-time shallowing of the light curve is not the turnover of a steep decay into a plateau of the canonical light curve, but a different transition at later times. This is an observed feature of GRB afterglows (e.g.,~\citealt{Evans2009, Li2012, Zaninoni2013}) and a change in circumburst medium is one of the few ways for the light curve of the blast wave to transition to a new slope.

From the light curves of simulations $\gamma5a4$, $\gamma10a100$, and $\gamma5a\frac{1}{100}$, as well as the light curves from the simulations not shown here ($\gamma5a100$ and $\gamma10a4$), we conclude that there is no rebrightening at times on the order of minutes to days for a blast wave encountering a change in circumburst medium. The simulations show that when the magnitude of the flux observed in the ISM is lower than the flux observed in the stellar wind environment at the time of the encounter (top left panel in Figure~\ref{fig:scaling}) the blast wave conforms to the new slope and there is no rebrightening. When the magnitude of the flux observed from a blast wave in an ISM environment is higher than the magnitude of the flux observed from a blast wave in the stellar wind environment, the blast wave still smoothly transitions to the new environment. In all cases, though, the flux observed from the blast wave after the encounter is lower than the flux observed from a blast wave that has only traveled in the ISM. 

To test whether the light curves for our simulations are an accurate depiction of a blast wave encountering a circumburst density change, or if they only represent the light curves for encounters at fluid Lorentz factors of $\gamma \leq 10$, with changes in density of factors $\leq 100$ and with $\theta_0=0.1$, we analytically solve for the light curves of various blast wave encounters for various encounter Lorentz factors, $\gamma_{enc}$, density change factors, and jet half opening angles, $\theta_0$. The top right panel in Figure~\ref{fig:scaling} shows light curves for one dimensional analytical models of blast waves traveling in a stellar wind environment that encounter a density change of factor 10 followed by a homogeneous medium with the encounters happening at varying Lorentz factors. When the factor by which the density changes is kept fixed, the shape of the light curve remains independent of the actual time of the encounter and light curves for different encounter times can simply be shifted to match, as can be seen from the figure. 

The bottom left panel in Figure~\ref{fig:scaling} shows the light curves for blast waves traveling in the stellar wind environment that encounter a density change of varying magnitudes at $\gamma_{enc}=25$. We show here density changes ranging from $10^{-8}$ to roughly correspond with the density drops used in \cite{Mesler2012}, to $10^3$. For larger density jumps, the blast wave quickly becomes non-relativistic, causing our analytical model to become less accurate. Lastly, the bottom right panel in Figure~\ref{fig:scaling} shows light curves for blast waves traveling in a stellar wind environment that encounter a density jump of factor 10 at $\gamma_{enc}=25$ for various jet half opening angles, $\theta_0$. The only light curve in this figure that has some resemblance of a flare is that of the blue line, which has an opening angle very close to $\gamma_{enc}\sim 1/\theta_0$, meaning that for any flare to possibly occur around the time of the encounter, $\gamma_{enc}=1/\theta_0$. This is the setup of our simulations $\gamma10a4$ and $\gamma10a100$: $\gamma_{enc}=10$, and $\theta_0=0.1$, and no flare is observed in one or two dimensions.

From this analysis, we conclude that for a blast wave traveling in a stellar wind environment that encounters a sudden change in circumburst environment, there is no observable rebrightening regardless of the size of the jump, drop, or fluid Lorentz factor at the time of the encounter that we have studied. There is a bump in the light curve at the observation angle of $\theta_{obs}=0.2$ radians for the light curve of simulation $\gamma10a100$ in 2D, but $\theta_{obs}>\theta_0$ and thus this is not a likely indication of an observable flare. We do not see a flare from the forward shock, nor do we see a flare caused by the reverse shock at times corresponding to the afterglow flares seen by {\it Swift}. 

We study a single spherically symmetric change in the circumburst environment. However, the actual environment may have a more complicated structure with instabilities (e.g., \citealt{vanMarle2012}) or thin circumburst shells~\citep{Mesler2012}. A study of asymmetric structures (i.e., density clumps) around the progenitor would require three dimensional simulations and is beyond the scope of this paper. We do look at the interaction with circumburst shells, and our results are discussed further in Appendix~\ref{App:AppendixB}. In this case too, we have been unable to reproduce the flares reported by \cite{Mesler2012}.

\subsection{Very Late Time Rebrightening}

The simulations of a density jump did show flares at times long after the encounter occurred (at times much later than plotted in Figure~\ref{fig:sim_lc}). At such late times (on the order of $10^4$ days), these flares are beyond X-ray observations, but may be of interest at radio frequencies.  There are two contributors to these flares--the fireball conditions created from the strong reverse shock that causes a secondary blast wave to form, shown in the bottom row of Figure~\ref{fig:sim_gamma}, and the encounter of the receding jet with the wind termination shock. We discuss the contribution of the secondary blast wave first, followed by a discussion of the receding jet.

\begin{figure*}
\includegraphics[width=0.49\textwidth]{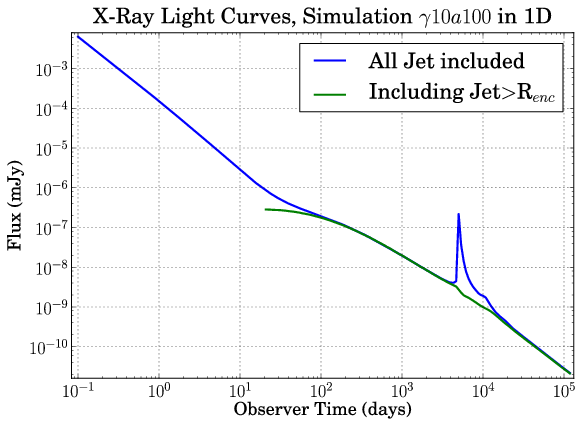}
\includegraphics[width=0.49\textwidth]{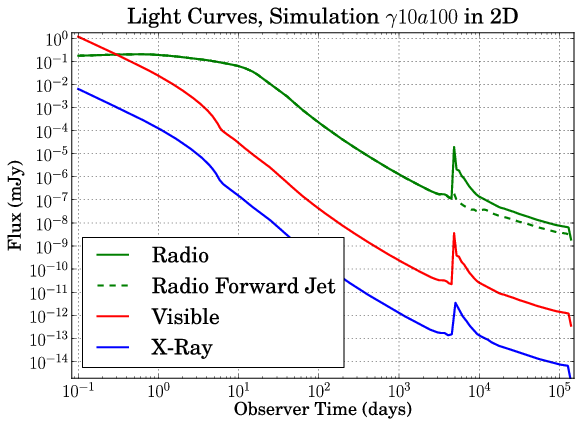}
\caption{Figures showing the late time flare of simulation $\gamma10a100$ ($\gamma_{enc}=10$, observed at roughly 3 days, with a jump of 100 and the jet half-opening angle of $\theta_0 = 0.1$). Left: light curve of the one dimensional simulation $\gamma10a100$ including the entire evolution (blue line) and including only the fluid at radii larger than $R_{enc}$ (green line).  Right:  light curves for the 2D simulation $\gamma10a100$ at three frequencies:  radio, $\nu =  4.86\times 10^9$ Hz, visible, $\nu = 4.56\times 10^{14}$ Hz, and X-ray, $\nu = 5\times 10^{17}$ Hz. These light curves were calculated on axis with the blast wave, $\theta_{obs} = 0.0$ radians. The blue dashed line shows the light curve calculated at a radio frequency using only the forward jet (and not including the receding jet). \label{fig:rebright}}
\end{figure*}

Figure~\ref{fig:rebright} shows the very late time flare observed from the one and two dimensional simulations of simulation $\gamma10a100$. Comparing the time at which the flare begins for simulation $\gamma10a100$ in 1D and 2D, shown in Figure~\ref{fig:rebright}, with the time and radius of the secondary blast wave shown in the bottom left panel in Figure~\ref{fig:sim_gamma}, it is apparent that the secondary blast wave is a cause of the rebrightening. This is confirmed by the left panel in Figure~\ref{fig:rebright}, which shows the light curve calculated for simulation $\gamma10a100$ in 1D using the entire blast wave (blue line), against the light curve observed from the parts of the blast wave at radii larger than the encounter radius (green line). The left panel in Figure~\ref{fig:rebright} clearly illustrates that the flare is caused by the fluid inside the encounter radius at late times, and the most energetic fluid in the 1D simulation that is at radius smaller than $R_{enc}$ at late times is the secondary blast wave. 

The left panel in Figure~\ref{fig:rebright}, although explaining the flare for the one dimensional case of simulation $\gamma10a100$, does not explain the small flares seen in the one dimensional conical outflow for simulation $\gamma5a4$ nor the flares for the two dimensional simulations $\gamma5a4$ and $\gamma10a100$.

To fully understand the cause of this late time flare, we must emphasize that it is seen at a wide range of observer angles and frequencies (right panel in Figure~\ref{fig:rebright}) at the same observer time, and is observable at the same time as the flare from the secondary blast wave. The secondary blast wave cannot be the cause of the flare for the 2D simulation because it is quickly slowed by the surrounding material (seen in the bottom right panel in Figure~\ref{fig:sim_gamma}), and is also not a characteristic of simulation $\gamma5a4$ in one or two dimensions. Moreover, the flare observed in simulation $\gamma5a4$ is observable in the 2D light curves and the 1D light curve that assumed conical outflow, but not in the 1D light curve assuming spherical outflow. 

The cause of the late time flare is the receding jet that encounters the sudden density jump (but not from a density drop). The reason this event seemingly occur at the same time as the secondary blast wave is because the sum of the distance traveled by the reverse shock and the flux from the secondary blast wave is the same as the distance traveled by the flux observed from the receding jet encountering the density jump. In short, the density jump results in a sudden increase in flux traveling in the opposite direction of the blast wave itself, meaning a spike in the light curve will not be observed at the time of the encounter because the light from the encounter is traveling away from the observer. This is confirmed by the green lines in the right panel of Figure~\ref{fig:rebright}, which shows that the light curve observed only from the forward jet does not result in a very late time flare whereas the light curve observed from both the forward and the receding jet does have a flare. We obtain the same results when analyzing simulation $\gamma10a100$ at different frequencies, and by analyzing the very late time flare of simulation $\gamma5a4$ in 2D (and 1D conical outflow). The reason for the differing behavior of the green line with the red and blue lines of the right panel of Figure~\ref{fig:rebright} is that the radio frequency corresponds to a different spectral regime than the visible and X-ray frequencies at early times. The radio frequency at early times is below the synchrotron peak critical frequency, $\nu_m$, meaning the light curve scales with $t^0$ in the wind scenario (e.g., \citealt{Granot2002}). This early time radio light curve behavior is showing a different spectral regime with $\nu_m$ passing through the observed band at the turnover.

The receding jet, or counter jet, has been known to be observable at very late times \citep{Granot2003, Li2004, Zhang2009, Wang2009}, with its magnitude being a function of the density structure into which the forward and receding jets are traveling~\citep{Wang2009}. What differentiates the flares from the receding jet shown in this paper and the visibility of the receding jet in previous work is that this flare is from the encounter of the receding jet with the sudden higher density region, as opposed to the counter jet being observable as it becomes non-relativistic and isotropic~\citep{Li2004} or as it travels into a differing circumburst medium than for the forward jet, or as both the forward and receding jet travel into a highly dense medium~\citep{Wang2009}. In addition, the flare in this paper is observable at a wide range of observer angles, in contrast to the prediction that the receding jet is only observed at optimal angles~\citep{Granot2003}. The very late time flare observed in this work is from the sudden decrease in Lorentz factor of the jets at the encounter, causing a flare to be observed as opposed to observing the receding jet itself.

From our simulations, the only flare calculated is that at much later times and much lower magnitudes in flux than those observed by {\it Swift}. The only possible flare on the time scale of observed flares we are able to observe with a sudden change in circumburst medium is the small bump in the light curve seen at around 30 days at $\theta_{obs}=0.2$ radians in the middle right panel in Figure~\ref{fig:sim_lc}, but this occurs only for $\theta_{obs}>\theta_0$ and thus is unlikely to be observed.

\section{Summary and Conclusions}\label{sec:SaC}

We have shown numerically and analytically that a blast wave evolving partially in a stellar wind environment that encounters a sudden change in density, either an increase or a decrease, followed by a constant density environment for a wide range of initial conditions does not cause an observable rebrightening. Flares at very late times are a function of the size of the density jump--the larger the density jump at the encounter, the stronger the reverse shock, resulting in a brighter late time flare. We studied light curves for a wide range of frequencies, and although this paper focuses on X-ray light curves, we saw similar behaviors for a wide range of frequencies. These flares potentially might be observable at radio frequencies under favorable conditions (Figure~\ref{fig:rebright}). This answers the first question listed in Section~\ref{sec:intro}: {\it the size of the density jump {\bf does} affect the dynamics and resulting light curve and may cause a faint very late time flare at radio frequencies for large jumps, but there are no observable flares at X-ray frequencies}. 

We found that for a blast wave traveling in a stellar wind environment encountering an ISM environment, the resulting flux observed will gradually transition from one environment to the next. As seen in Figure~\ref{fig:scaling}, if the flux observed from a blast wave traveling solely in the ISM is lower than the flux observed with a blast wave traveling solely in the wind environment at the time of the encounter, the flux observed from the blast wave will simply dim and follow the same light curve slope of the ISM. If the flux observed from a blast wave traveling solely in the ISM is higher than the flux observed from a blast wave traveling solely in a wind environment at the time of the encounter, the observed flux from the blast wave will not suddenly increase, but stay relatively steady as it transitions to the new slope of the ISM. This analysis was mostly done for the one dimensional simulations and showed that the light curve is affected by the size of the density change, but not the Lorentz factor at the encounter. We have explored various ratios of $\gamma_{enc}/\gamma_{jet}$ with two dimensional simulations and found that this conclusion holds in two dimensions. However, strictly speaking the recollimation seen in the two dimensional simulations of density jumps is dependent on the Lorentz factor of the fluid at the encounter, and the amount by which the blast wave recollimates should be inversely proportional to the fluid encounter Lorentz factor (i.e., blast wave recollimates more for lower fluid encounter Lorentz factors), although this is not expected to have an effect on the occurrence of a flare.

We have studied the two dimensional and one dimensional effects of a density drop and have shown that the blast wave {\it does increase in speed, but does {\bf not} recollimate}. There is also {\it no rebrightening} caused by this sudden increase in blast wave speed (Figure~\ref{fig:sim_lc}). Our two dimensional studies of a density jump have yielded the conclusion that {\it there is some sideways spreading from a high energy collision} of the blast wave with a large jump in circumburst density, and the amount by which the blast wave spreads is {\it highly dependent on the size of the density jump}. There is also a flare at very late times, but this is not caused by the sideways spreading. The late time flare is caused from {\it the reverse shock of the receding jet} (Figure~\ref{fig:rebright}). This very late time flare only occurs years after the initial flux observed from the blast wave, and data from {\it Swift} is only gathered for days after the first flux is observed resulting in the conclusion that this is not the flare observed by {\it Swift}. In addition, these flares range from being $10^{-4}$ to $10^{-10}$ orders of magnitude smaller than the initial flux of the blast wave which is below the threshold of {\it Swift}.

We have analytically modeled the light curves of one dimensional simulations, and are able to use the model for on axis two dimensional dynamics that do not cause strong reverse shocks as well. From our analytical solutions and our numerical models, we have answered the questions listed in the Introduction and have concluded that a blast wave traveling in a stellar wind environment that encounters a change in density followed by an ISM environment will not cause observable flares.

Our work confirms earlier studies (NG, \citealt{vanEerten2009windtermination}) that a wind termination shock will not cause flares in light curves and extends this analysis to density drops and to two dimensional simulations. We were unable to reproduce the flares seen in PW due to the discrepancy of the fluid Lorentz factor during the encounter--the analysis done in PW assumed a constant fluid Lorentz factor during the encounter, and we have extended this analysis to more accurately model the decrease in fluid Lorentz factor at the encounter. We were not able to reproduce the flares seen in \cite{Mesler2012} and \cite{Eldridge2006}. Both of these studies use an analytical model for the blast wave based on total swept up mass. They do not account for the radial or angular structures of the blast wave, such as the differing shock regions shown in Figure~\ref{fig:shocks} and the spreading of the shock as it becomes non-relativistic. Without accounting for these radial and angular structures, their models of the change in Lorentz factor with change in mass result in a discrepancy in time scale and flux observed between ours and their results. In addition, as the blast wave becomes non relativistic, the blast wave spreads outwards and its width becomes comparable to the radius of the blast wave. Thus, for the the interaction of a blast wave with a spike, like that in~\cite{Mesler2012}, the swept up mass model does not allow for capturing features on the same or shorter time scale of the blast wave crossing the shell because the width of the shell is smaller than the width of the blast wave. This is discussed more thoroughly in Appendix~\ref{App:AppendixB}.

The very late time radio flares from our simulations are similar to those seen in simulations of multiple shell models~\citep{Maxham2009, Vlasis2011}. However, the simulations of the multiple shell model are simulated by adding energy to the system at late times leading to a different light curve slope after the encounter. The very late time radio flares seen in this paper were calculated from a simulation in which no energy was added, resulting in the flares seen in this paper being of different origin than from the multiple shell models.

We conclude that a wind termination shock, or more generally, any sudden transition in circumburst density (even extreme changes), is very unlikely to be the cause of the flares observed by {\it Swift} because those flares return to the same baseline as the light curve prior to the flare \citep{burrows2005}. With the encounter of a change in circumburst environment, the only observable flares do not return to the same baseline because flux from the receding jet is also being observed. Moreover, if the flare occurred at the time of the encounter, the light curve would transition to the new slope of the ISM material, {\it not} the slope of the stellar wind. This makes it highly unlikely that a flare seen at the encounter with the change in circumburst environment could be the explanation for the flares seen by {\it Swift}.

Through our numerical and analytical analysis of a blast wave encountering a circumburst density change that originated from the same point, we have concluded that a flare will not be observed by {\it Swift}, but a change in environment could be the explanation for the late-time shallowing observed for some GRB afterglows \citep{Evans2009, Li2012, Zaninoni2013}. It would be of interest to compare our results with observational data in future research.

\acknowledgments

This research was supported in part by NASA through grant NNX10AF62G issued through the Astrophysics Theory Program and by the NSF through grant AST-1009863 and by the {\it Chandra} grant TM3-14005X. Resources supporting this work were provided by the NASA High-End Computing (HEC) Program through the NASA Advanced Supercomputing (NAS) Division at Ames Research Center. The software used in this work was in part developed by the DOE-supported ASCI/Alliance Center for Astrophysical Thermonuclear Flashes at the University of Chicago. We wish to thank Ehud Nakar for helpful discussions.\\

\appendix
\section{Grid Setup and Resolution}\label{App:AppendixA}
To run our simulations, we first found the radius and time corresponding to the Lorentz factor at which we wanted the external density change to occur, the size of the grid, as well as the needed resolution of the simulation. We use Equations~\ref{eq:r0} and~\ref{eq:initial_time} to set up the initial shock location and time respectively with the initial fluid Lorentz factor of 15. Next, we use Equation~\ref{eq:renc} to find the location of the Lorentz factor at which the jump occurs, and a form of Equation~\ref{eq:gamma_enc} to calculate the time at which the Lorentz factor is equal to that at the jump.

The grid size needs to be large enough to ensure the simulation would run a satisfactory amount of time after the encounter. We choose a maximum observation time, $t_{obs}$, for the simulation to run, which gives an equation for the maximum radius and simulation time:
\begin{equation}\label{eq:tobs}
t_{obs} = t - \frac{R}{c}.
\end{equation}

$R$ in Equation~\ref{eq:tobs} is the maximum size of the grid, which we will refer to as $r_{max}$, and can be represented as the sum of the radius of the encounter, $R_{enc}$, plus some distance covered, $\Delta r$.
\begin{equation}\label{eq:rmax}
R = r_{max} = R_{enc} + \Delta r
\end{equation}
The variable $\Delta r$ in Equation~\ref{eq:rmax} can be found by integrating the velocity over time. We use a much simplified model for the grid setup because we need only a rough approximation. Using this simplified model, we assume that immediately after the encounter, the blast wave returns to a BM self similar solution of a blast wave in a homogeneous environment. This, again, is not an accurate assumption, but accurate enough for this purpose. 

Velocity is represented by $v = c\beta$ and $\beta = \sqrt{1-\frac{1}{\Gamma_{2}}}$ where the subscript of $2$ here denotes that this is the Lorentz factor after the encounter. $\Gamma_2$ if found by a similar equation to Equation~\ref{eq:gamma_enc} except with $k=0$.
\begin{equation}
\Gamma_2 = \Gamma_{enc}\left(\frac{t_2}{t_{enc}}\right)^{\frac{-3}{2}}
\end{equation}
Using the above equations, we obtain the integral of velocity over time.
\begin{equation}\label{eq:integral}
\Delta r= \int_{t_{enc}}^{t_f} \left[1 - \frac{1}{\Gamma_{enc}^2}\left(\frac{t_2}{t_{enc}}\right)^3\right]^{\frac{1}{2}}dt.
\end{equation}

To solve Equation~\ref{eq:integral}, we can expand the integrand.
\begin{equation}\label{eq:expansion}
\left[1 - \frac{1}{\Gamma_{enc}^2}\left(\frac{t_2}{t_{enc}}\right)^3\right]^{\frac{1}{2}} = 1 - \frac{1}{2\Gamma_{enc}^2}\left(\frac{t_2}{t_{enc}}\right)^3 - \frac{1}{8\Gamma_{enc}^4}\left(\frac{t_2}{t_{enc}}\right)^6
\end{equation}

Using Equation~\ref{eq:expansion} in Equation~\ref{eq:integral} to find $\Delta r$, we derive $r_{max}$ from Equation~\ref{eq:rmax}. Next, we calculate the corresponding simulation time at the end of the grid using Equation~\ref{eq:tobs}.

We set the minimum value of the grid to a size that ensures the entire initial shock is shown, $r_{min} = 0.01R_0$.

Lastly, to find the optimal resolution, we need to be certain that the initial shock is resolved. We use adaptive mesh refinement and we specify the number of base blocks, $\mathrm{bb}_r$, the number of cells per block, cpb, and the maximum refinement level, ref,  for the simulation. Equation~\ref{eq:refinement} describes the relationship these values have to the resolution where $A$ is the amount $\Delta R$ is refined:
\begin{equation}\label{eq:refinement}
\frac{r_{max} - r_{min}}{\mathrm{bb}_x\times \mathrm{cpb}\times2^{\mathrm{ref} - 1}} = \frac{\Delta R}{A},
\end{equation}
\begin{figure}
\begin{center}
\includegraphics[width=0.49\textwidth]{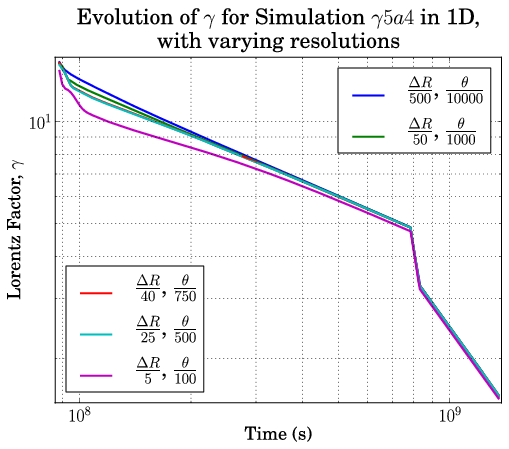}
\caption{Plot of the Lorentz factor over time for simulation $\gamma5a4$ (blast wave encountering a density jump of factor 4 at a time and radius corresponding to the front of the shock fluid Lorentz factor, $\gamma_{enc}=5$) in 1D for various resolutions. \label{fig:setup1_resolution_test}}
\end{center}
\end{figure}
$\Delta R = \frac{R_0}{12\Gamma_0^2}$,  cpb = 8, and the optimal value of $A$ can be tested, as shown in Figure~\ref{fig:setup1_resolution_test}. Figure~\ref{fig:setup1_resolution_test} shows that even with a very small amount of resolution in the radial direction, the simulation still converges at the time of the encounter--the part of the simulation on which we are focusing. The refinement level in the $\theta$ direction must also be computed. The simulation only computes from $0\leq\theta\leq \frac{\pi}{2}$, which yields:
\begin{equation}\label{eq:theta_refine}
\frac{\pi/2}{\mathrm{bb}_{\theta}\times \mathrm{cpb} \times 2^{\mathrm{ref}-1} }= \frac{\theta_0}{B}
\end{equation}
where bb$_{\theta}$ is the number of base blocks in the $\theta$ direction, $\theta_0$ is the jet half opening angle, and $B$ is the number by which we want the width of the jet opening to be refined. 

Using the equations in this section, we are able to set up the simulation grid for optimal results.

\section{Circumstellar Shells}\label{App:AppendixB}

It has been proposed that circumstellar shells could cause flares in light curves~\citep{Mesler2012}. We test this specific scenario of the interaction of a blast wave with a narrow shell in which the blast wave becomes non-relativistic. We model this scenario as a stellar wind environment followed a large density jump of small width that is terminated by a large density drop followed by a homogeneous medium. We model the setup of~\cite{Mesler2012} but exclude the wind termination shock for we have analyzed the effect of a wind termination shock in this paper and have concluded that will not cause a flare. Although we do not model the wind reverse shock in this scenario, the reverse shock acts on the mass ejected by the stellar wind. The Lorentz factor at the encounter with the strong density jump, according to the swept up mass approximation, is therefore identical with or without accounting for the wind reverse shock. Our stellar wind environment setup is: 
\begin{equation*}
E_{iso}=10^{53}\;\; \mathrm{ergs,}\;\; \dot{m}_{\mathrm{wind}}=10^{-5}\Msun\mathrm{yr}^{-1}\mathrm{,}\;\; v_{\mathrm{wind}} = 10^3\;\; \mathrm{km/s,}\;\; \mathrm{and}\;\; \rho_{ref} = \frac{\dot{m}_{\mathrm{wind}}}{4\pi R_{ref}^2v_{\mathrm{wind}} }.
\end{equation*}
The blast wave then encounters the spike at $\gamma_{enc}=10$, and the density of the spike is $10^5\rho_{ext}(R_{enc})$, where $\rho_{ext}$ is that of Equation~\ref{eq:external_density}. The spike contains the amount of mass $\mathrm{M}_{\mathrm{spike}} = 0.1\Msun$, and is followed by a sudden lower density region of $10^{-7}\rho_{ext}(R_{enc})$. This is essentially the same setup as that of~\cite{Mesler2012} that resulted in light curve flares in their analysis.

\begin{figure*}
\includegraphics[width=0.49\textwidth]{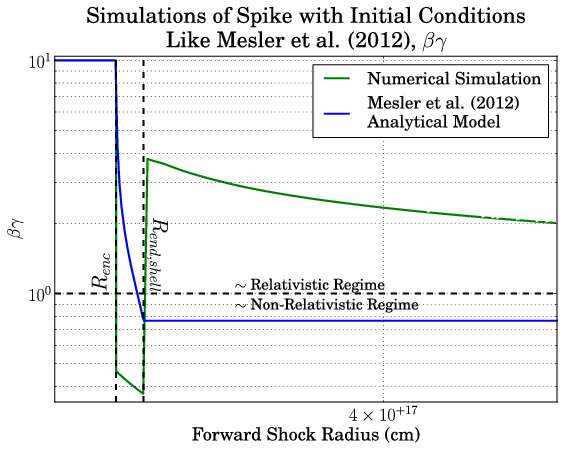}
\includegraphics[width=0.49\textwidth]{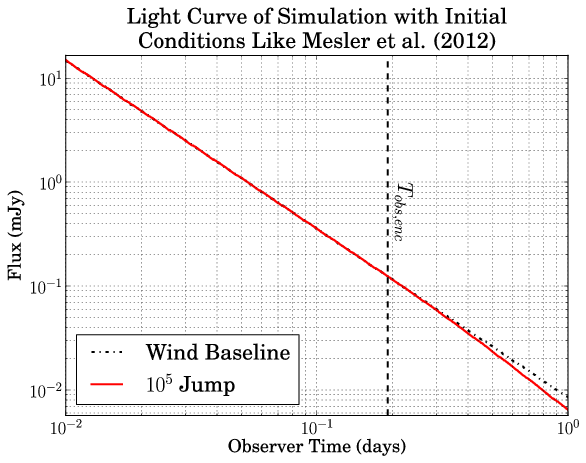}
\caption{Figures for the setup with a spike in the circumburst density profile. The spike is of width $3.15\times10^{14}$ cm at a radius corresponding to $\gamma_{enc}=10$, with an initial density jump of factor $10^5\rho_{ext}(R_{enc})$, followed by a drop of $10^{-7}\rho_{ext}(R_{enc})$. Left: plot of the forward shock velocity in terms of $\beta\gamma$ over radius for our numerical simulation and the analytical model of~\cite{Mesler2012}. Right: X-ray light curve calculated from our numerical simulation of this setup. \label{fig:spike}}
\end{figure*}

We numerically simulate this setup with a resolution in time of roughly 30 snapshots during the encounter to ensure the behavior of the blast wave within the spike is resolved, and no flare is observed (right panel in Figure~\ref{fig:spike}). The discrepancy between our results and that of~\cite{Mesler2012} is that~\cite{Mesler2012} models only swept up mass and does not account for the radial structure of the blast wave. With this approximation, the mass swept up after the encounter instantaneously mixes with the mass swept up prior to the encounter (which are actually separated by the contact discontinuity) resulting in an overestimation of the total mass radiating with the post-encounter Lorentz factor. In addition to the over-estimation of the mass in the forward shock region, this method also overestimates the Lorentz factor during the encounter. To show this, we analytically modeled the Lorentz factor evolution as done in~\cite{Mesler2012} (who follow~\citealt{Peer2012}), and compared it to the numerical results for the same setup. This is shown in the left panel in Figure~\ref{fig:spike}. This figure depicts the overestimation of the Lorentz factor from the assumptions of~\cite{Mesler2012}, and the underestimation of the Lorentz factor in the homogeneous medium. The Lorentz factor jumps up immediately after exiting the spike, but there is still no flare from this because initially, only negligible mass is radiating with this larger Lorentz factor. The light curve from our numerical simulation is plotted in the right panel in Figure~\ref{fig:spike}, and shows a steepening in post-encounter slope but does not show a flare. No flare is observed because the fluid behind the contact discontinuity is still traveling at the pre-encounter Lorentz factor immediately after the FS hits spike. The FS Lorentz factor immediately drops, resulting in it only minimally contributing to the total flux observed. In the swept up mass approximation, the FS Lorentz factor gradually lowers resulting in it still contributing noticeable amounts to the light curve immediately after the encounter. This may be the cause of the flares seen in~\cite{Mesler2012}, however, when calculating the emission from a spatially resolved blast wave, while taking light travel times into account from different radii and angles, we find that no observable flare is produced.

The light curve shown here is calculated at the X-ray frequency of $5\times10^{17}$ Hz without the contributions from electron cooling or absorption. However, we did calculate light curves with the contributions from electron cooling and absorption, and saw no qualitative differences. We also considered light curves for different frequencies such as optical and radio. Light curves calculated at optical frequencies exhibited the same behavior as the X-ray light curves.  Below the synchrotron self-absorption break a steep drop was observed following the onset of the shell encounter and a steep rise when the blast wave emerged from the shell. However, the (thin) emitting region of the blast wave in the optically thick regime was not always fully resolved numerically, leading to noise in the light curves. Regardless, our conclusions regarding X-ray light curves, which are the main focus of our study, remain unaffected.

We believe that any model where the blast wave itself is not resolved is fundamentally unsuited to describe features occurring on time scales $\Delta R / c$. In the non-relativistic case, the relevant light crossing time is of the order of the blast wave radius, $R / c$, due to both the absence of relativistic beaming of the emission and the width of the blast wave $\Delta R$ becoming a sizable fraction of the radius.

\bibliography{windtermination}

\end{document}